\documentclass[12pt]{article}
\usepackage{amssymb,amsmath}
\usepackage{bm} 
\usepackage{booktabs} 
\usepackage{array}
\usepackage{latexsym}
\usepackage{graphicx}
\usepackage{color}
\usepackage{slashed}
\usepackage{datetime}
\usepackage[nosort]{cite}
\usepackage{verbatim}
\usepackage{enumerate}
\usepackage{chngpage} 
\allowdisplaybreaks
\usepackage{psfrag}
\usepackage{mathtools}
\usepackage{mciteplus}
\usepackage{dsfont}
\usepackage{multirow} 
\usepackage{physics}

\usepackage[colorlinks=true,      linkcolor=blue,      urlcolor=blue,      
            filecolor=blue,      citecolor=blue,       pdfstartview=FitH,      
						pdfpagemode=UseNone,      bookmarksopen=true]{hyperref}  
\usepackage[all]{hypcap}     

\def\({\left(}
\def\){\right)}
\def\[{\left[}
\def\]{\right]}

\def\One{{\hbox{ 1\kern-.8mm l}}}

\def\barray{\begin{array}}
\def\earray{\end{array}}
\def\be{\begin{equation}}
\def\ee{\end{equation}}
\def\bea{\begin{eqnarray}}
\def\eea{\end{eqnarray}}
\def\bal{\begin{align}}
\def\eal{\end{align}}

\numberwithin{equation}{section} 
\allowdisplaybreaks  

\makeatletter
\g@addto@macro\bfseries{\boldmath}
\makeatother

\definecolor{cardinal}{rgb}{0.6,0,0}
\definecolor{darkgreen}{rgb}{0,0.4,0}
\definecolor{golden}{rgb}{0.92, 0.7, 0}
\definecolor{midnight}{rgb}{0, 0, 0.5}
\definecolor{darkblue}{rgb}{0, 0, 0.7}
\definecolor{purple}{rgb}{0.5, 0, 0.5}
\newcommand{\Red}{\color{red}}

\newcommand{\DarkBlue}{\color{darkblue}}

\def\DB{\DarkBlue}

\def\IB#1{{\bf \Red IB:} {\DB #1}}

\def\YL#1{{\bf \Red YL:} {\Purple #1}}

\usepackage{epsf}
\usepackage{cite}
\usepackage{fancyhdr}
\usepackage{bm}
\usepackage{enumerate} 

\usepackage{empheq} \empheqset{box=\fbox}
\usepackage[usenames,dvipsnames]{xcolor}

\newcommand{\MyRed}{\color [rgb]{0.8,0,0}}

\def\YL#1{{\MyRed [YL: #1]}}

\numberwithin{equation}{section}  
\allowdisplaybreaks

\usepackage{graphicx}
\usepackage{tikz}
\usetikzlibrary{decorations.markings}
\tikzset{->-/.style={decoration={
			markings,
			mark=at position #1 with {\arrow{stealth}}},postaction={decorate}}}
\usepackage{adjustbox}
\usepackage{pgfplots}
\pgfplotsset{compat=1.11}
\usepgfplotslibrary{fillbetween}
\usetikzlibrary{intersections}
\usetikzlibrary{patterns}

\pgfdeclarelayer{bg}
\pgfsetlayers{bg,main}

\begin{document}


\begin{flushright}
		\hfill{MPP-2022-136}
\end{flushright}
\vspace{35pt}

\vspace{3mm}

\begin{center}

{\huge {\bf The
(amazing)
 Super-Maze}}

\vspace{14mm}

{\large
\textsc{Iosif Bena$^a$, Shaun D.~Hampton$^a$, Anthony Houppe$^a$, \\ Yixuan Li$^{b,a}$, Dimitrios Toulikas$^a$}}
\vspace{12mm}

\textit{$^a$	 Institut de Physique Th\'eorique, \\  
Universit\'e Paris-Saclay, CNRS, CEA,\\
	Orme des Merisiers, Gif-sur-Yvette, 91191 CEDEX, France  \\}  
\medskip

\textit{$^b$	 Max–Planck–Institut f\"ur Physik, Werner–Heisenberg–Institut, \\
	F\"ohringer Ring 6, 80805 M\"unchen, Germany  \\}  
\medskip

\vspace{4mm} 
%

{\footnotesize\upshape\ttfamily  iosif.bena, shaun.hampton, anthony.houppe, dimitrios.toulikas  @ ipht.fr;\\ yixuan @ mpp.mpg.de} \\
\vspace{13mm}
 

\end{center}

\begin{adjustwidth}{10mm}{10mm} 

\begin{abstract}
\vspace{1mm}
\noindent

The entropy of the three-charge NS5-F1-P black hole in Type IIA string theory comes from the breaking of $N_1$ F1 strings into $N_1 N_5$ little strings, which become independent momentum carriers. In M theory, the little strings correspond to strips of M2 brane that connect pairs of parallel M5 branes separated along the M-theory direction. We show that if one takes into account the backreaction of the M-theory little strings on the M5 branes one obtains a maze-like structure, to which one can add momentum waves. We also show that adding momentum waves to the little strings gives rise to a momentum-carrying  brane configuration -- a super-maze -- which locally preserves 16 supercharges. We therefore expect the backreaction of the super-maze to give rise to a new class of horizonless black-hole microstate solutions, which preserve the rotational symmetry of the black-hole horizon and carry $\sqrt{5/6}$ of its entropy.

\end{abstract}
\end{adjustwidth}

\thispagestyle{empty}
\clearpage



\baselineskip=14.5pt
\parskip=3pt

\tableofcontents

\baselineskip=15pt
\parskip=3pt

\section{Introduction}

String Theory is famous for its ability to count the degrees of freedom that give rise to the entropy of supersymmetric black holes, but this counting is always done in a regime of parameters where the interactions are weak and the classical black-hole solution does not exist \cite{Strominger:1996sh,Maldacena:1996ky,Dijkgraaf:1996cv}.  This leaves open the question of how these microscopic degrees of freedom look like in the regime of parameters where the classical black hole solution exists. This question is hard to answer, because it is difficult to track these degrees as one moves to this regime of parameters, and also because the black-hole microscopic degrees of freedom look different in different duality frames.\footnote{For example, the entropy of the D1-D5-P black hole for example comes from 1-5 strings carrying fractional momentum quanta, while the entropy of the U-dual IIA F1-NS5-P black hole comes from fractionated little strings carrying integer momenta.}

Historically, the quest for understanding the black hole degrees of freedom has been pursued from the opposite direction: people have first constructed solutions with black-hole charges that do not have a horizon and that exist in the same regime of parameters as the classical black hole solution \cite{Giusto:2004kj,Bena:2005va,Berglund:2005vb,Bena:2006is,Bena:2006kb,Bena:2007kg,Bena:2007qc,Bena:2008wt,Bena:2010gg,Bena:2011dd, Bena:2015bea,Bena:2016ypk,Bena:2017geu,Bena:2017xbt,Bena:2018mpb,Ceplak:2018pws,Heidmann:2019zws,Heidmann:2019xrd,Mayerson:2020tcl, Shigemori:2020yuo, Bena:2020yii,Bena:2020iyw,Giusto:2020mup,Houppe:2020oqp,Ganchev:2021pgs, Ganchev:2021iwy,Bianchi:2016bgx, Bianchi:2017bxl, Heidmann:2017cxt, Bena:2017fvm, Avila:2017pwi,Tyukov:2018ypq}. By construction, these solutions (known as microstate geometries) describe {\em some} of the black-hole microstates. The second step was to use holographic tools to relate some of them to the corresponding states in the CFT that counts the black-hole entropy \cite{Kanitscheider:2006zf,Kanitscheider:2007wq,Taylor:2007hs,Giusto:2015dfa,Bombini:2017sge,Giusto:2019qig, Tormo:2019yus, Bena:2019azk, Rawash:2021pik, Ganchev:2021ewa}.  This endeavor has been very successful, and has produced some of the finest checks of the power of holography to date. However, despite the tremendous amount of geometries constructed\footnote{Of order $e^{\sqrt{N_1 N_5 N_p^{1/2}}}$ for the D1-D5 system \cite{Shigemori:2019orj,Mayerson:2020acj}.}, it has not been possible to construct geometries dual to the states that account for the total black-hole entropy.

In this paper we take the opposite approach: We start from a weakly-coupled system whose degrees of freedom count the entropy of the black hole, and attempt to track these degrees of freedom to the regime of parameters where the classical black-hole solution exists. Our starting point is the F1-NS5-P black hole in Type IIA string theory, whose entropy comes from the fractionation of each of the $N_1$ fundamental strings into $N_5$ \textit{little strings} living in the worldvolume of the NS5-branes \cite{Seiberg:1997zk,Kutasov:2001uf}. The resulting $N_1 \times N_5$ little strings wrap the common F1-NS5 direction and can carry momentum along this direction by transverse oscillations in the other four internal directions of the NS5 branes \cite{Dijkgraaf:1996cv}\footnote{In the rest of this paper, we refer to these microstates as ``Dijkgraaf-Verlinde-Verlinde (DVV) microstates'' or ``little-string microstates." It is important to note that the DVV microscopic counting is different in spirit from the Strominger-Vafa/D1-D5 counting: here the momentum is carried by $N_1 N_5$ fractionated strings carrying integer momenta, while in the D1-D5 CFT we have a single long effective string with modes with momentum quantized in units of $1/N_1 N_5$. Hence, the entropy of the DVV microstates comes from fractionating the F1 strings, while the Strominger-Vafa entropy comes from fractionating the momentum carriers.}. It is not hard to see that in the Cardy limit the entropy of these oscillations and of their fermionic superpartners is $S_{\rm little~strings} = 2 \pi \sqrt{(4+2) {N_1 N_5 N_p \over 6}}$, reproducing precisely the entropy of the F1-NS5-P black hole.

\begin{figure}[h]
\begin{center}
 \includegraphics[width=.94\linewidth]{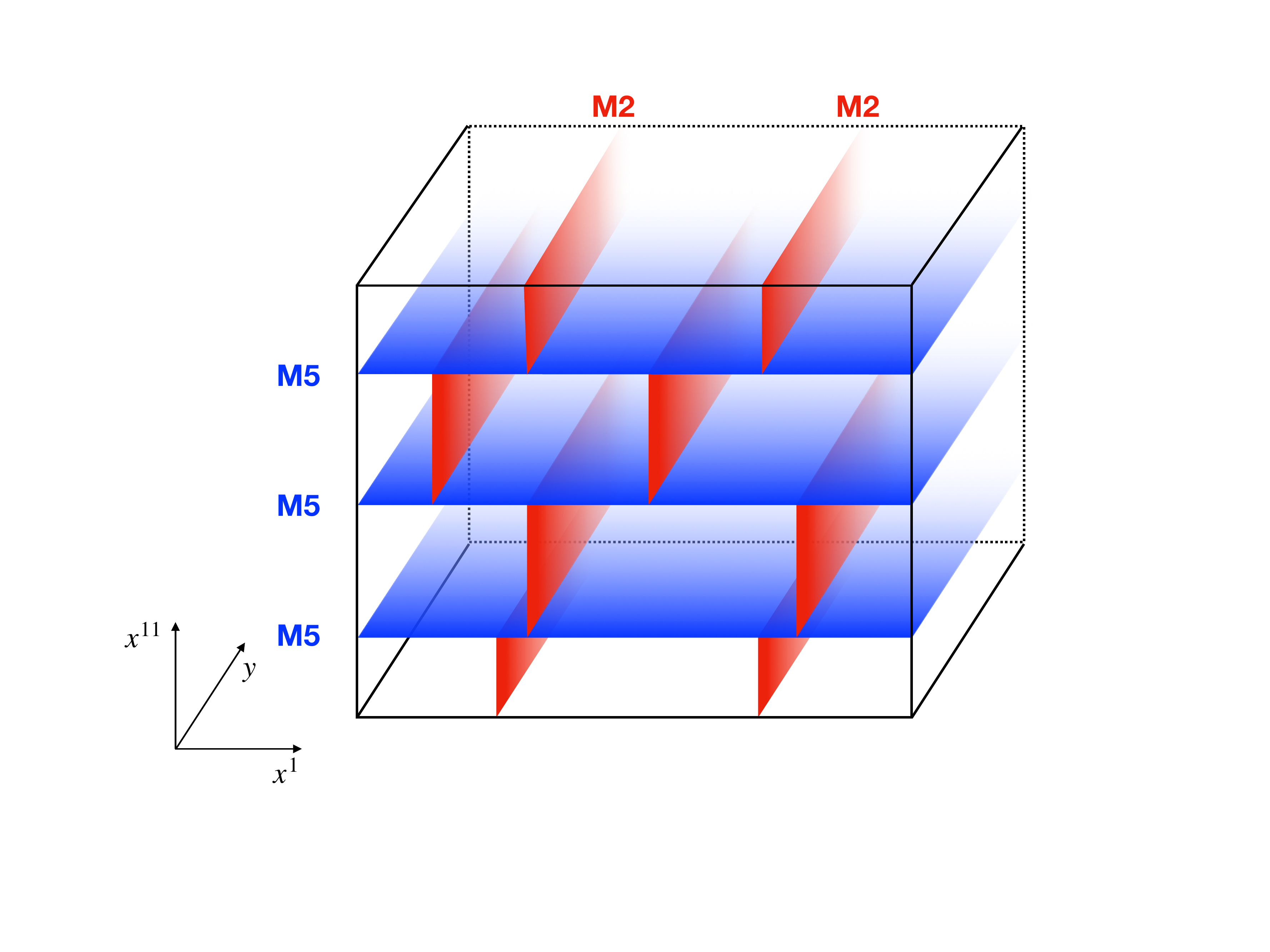}
\caption{Cross-section of $N_1=2$ M2 branes splitting into strips between $N_5=3$ M5 branes. The vertical axis is the M-theory direction, and the horizontal axis represents one of the internal directions of the M5 branes, $x^1$. The strips can carry momentum along the $y$-circle, which is common to the M2 and M5 branes.}
\label{fig:M2_fractionated_M5}
\end{center}
\end{figure}

The M-theory uplift of this system makes the little strings less mysterious: One obtains $N_5$ M5 branes wrapping the common 1-5 direction (that we will henceforth call $y$) and that are located at different points of the M-theory circle. One fundamental string uplifts to an M2 brane wrapping $y$ and the M-theory circle, $x^{11}$, and this M2 brane can break into $N_5$ ``strips'' stretching between two adjacent M5 branes. 
As one can see from Figure~\ref{fig:M2_fractionated_M5}, these M2 strips can move independently along the other internal directions of the M5 branes. Hence, the fractionation of an F1 string into $N_5$ little strings has a clear geometric  picture in M-theory, as the breaking of an M2 brane into $N_5$ strips.

The purpose of this paper is to begin tracking the fractionated little strings, from the ``zero backreaction'' regime, where their counting reproduces the the F1-NS5-P black-hole entropy, to the regime of parameters where their backreaction becomes important. We will show that the momentum-carrying fractionated strings coalesce into 4-supercharge brane bound-states that have locally 16 supersymmetries.

\begin{figure}[h]
	\centering
	\includegraphics[width=.31 \textwidth ]{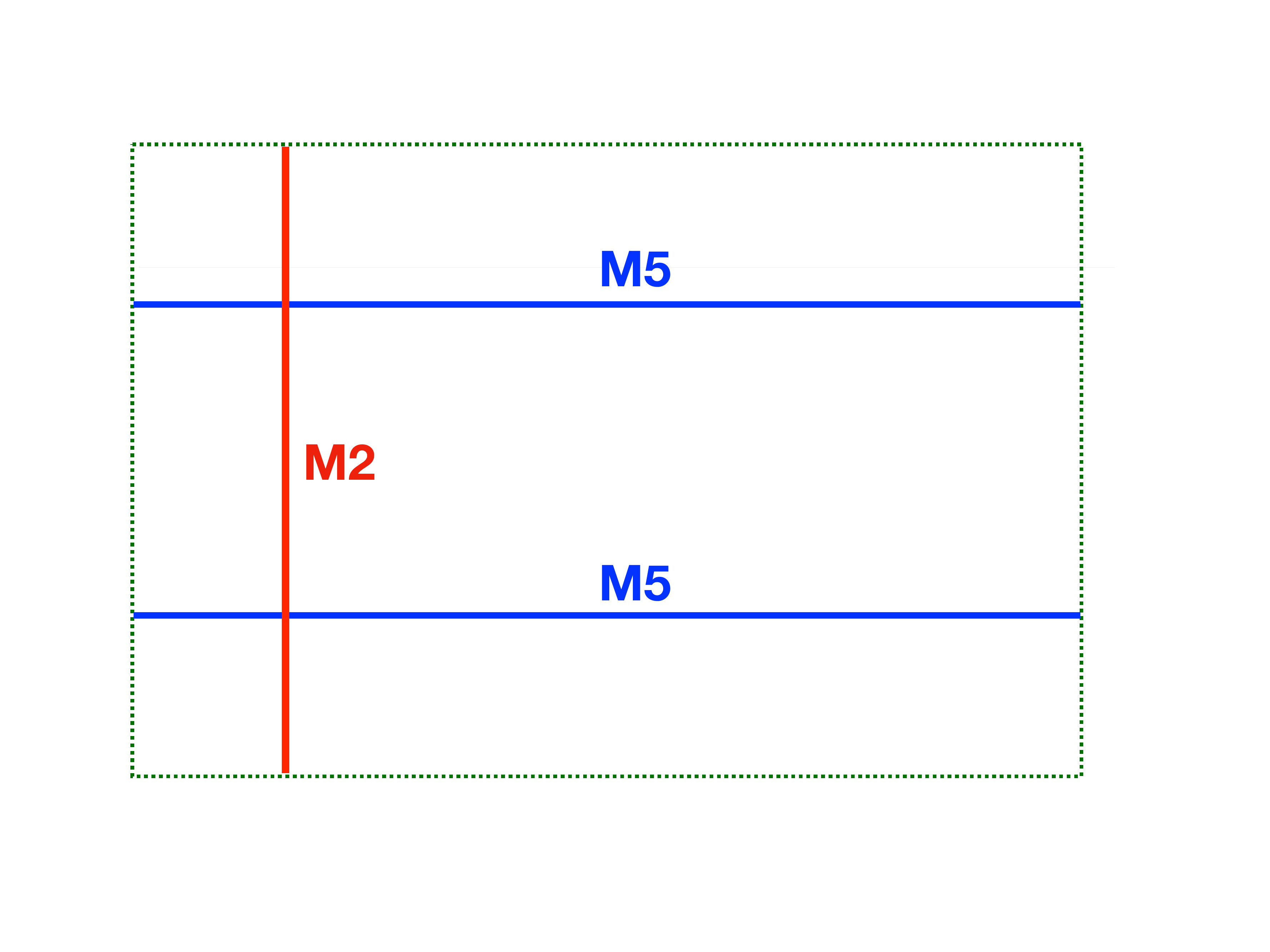}
	\includegraphics[width=.31 \textwidth]{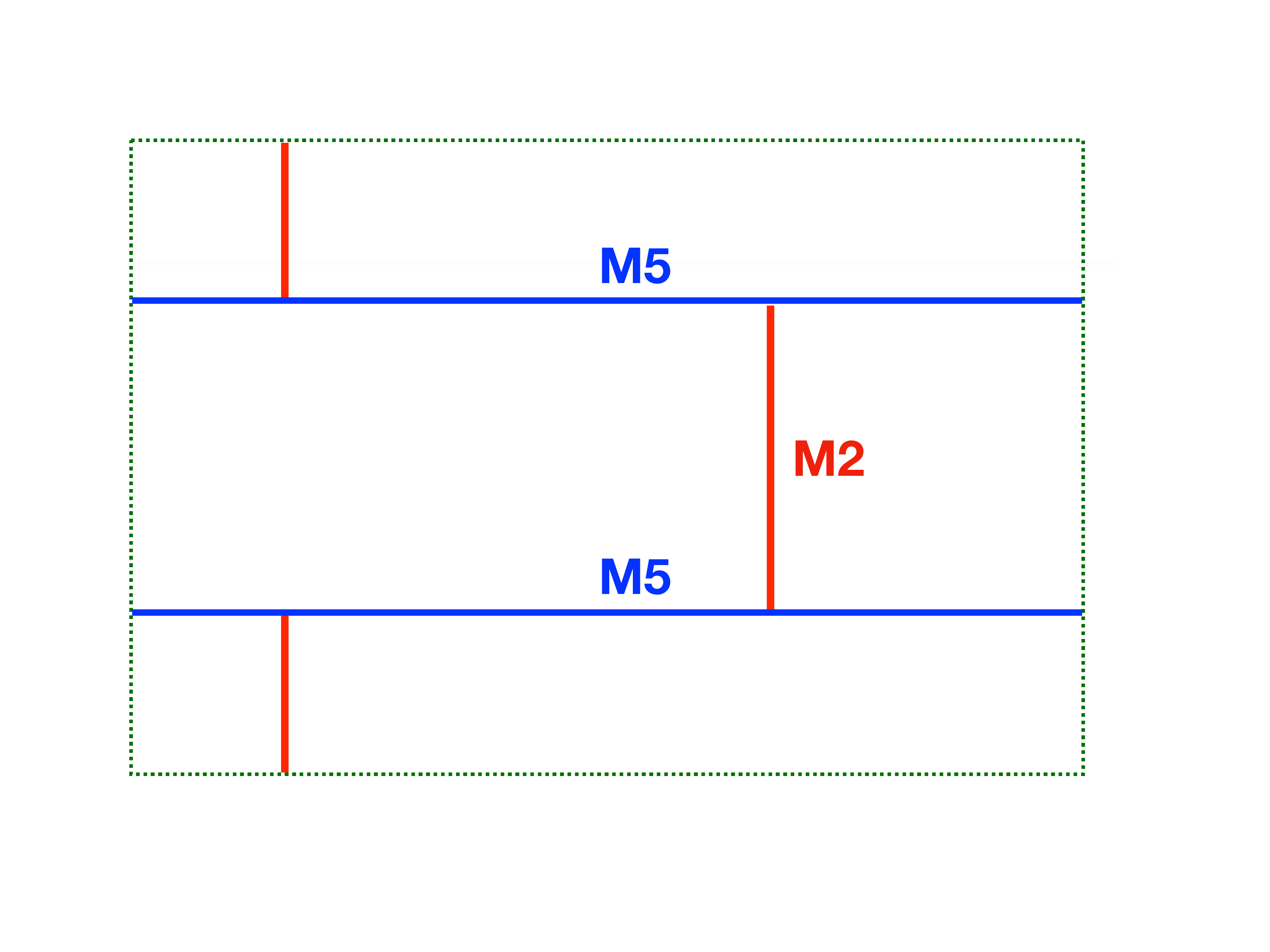}
	\includegraphics[width=.31 \textwidth]{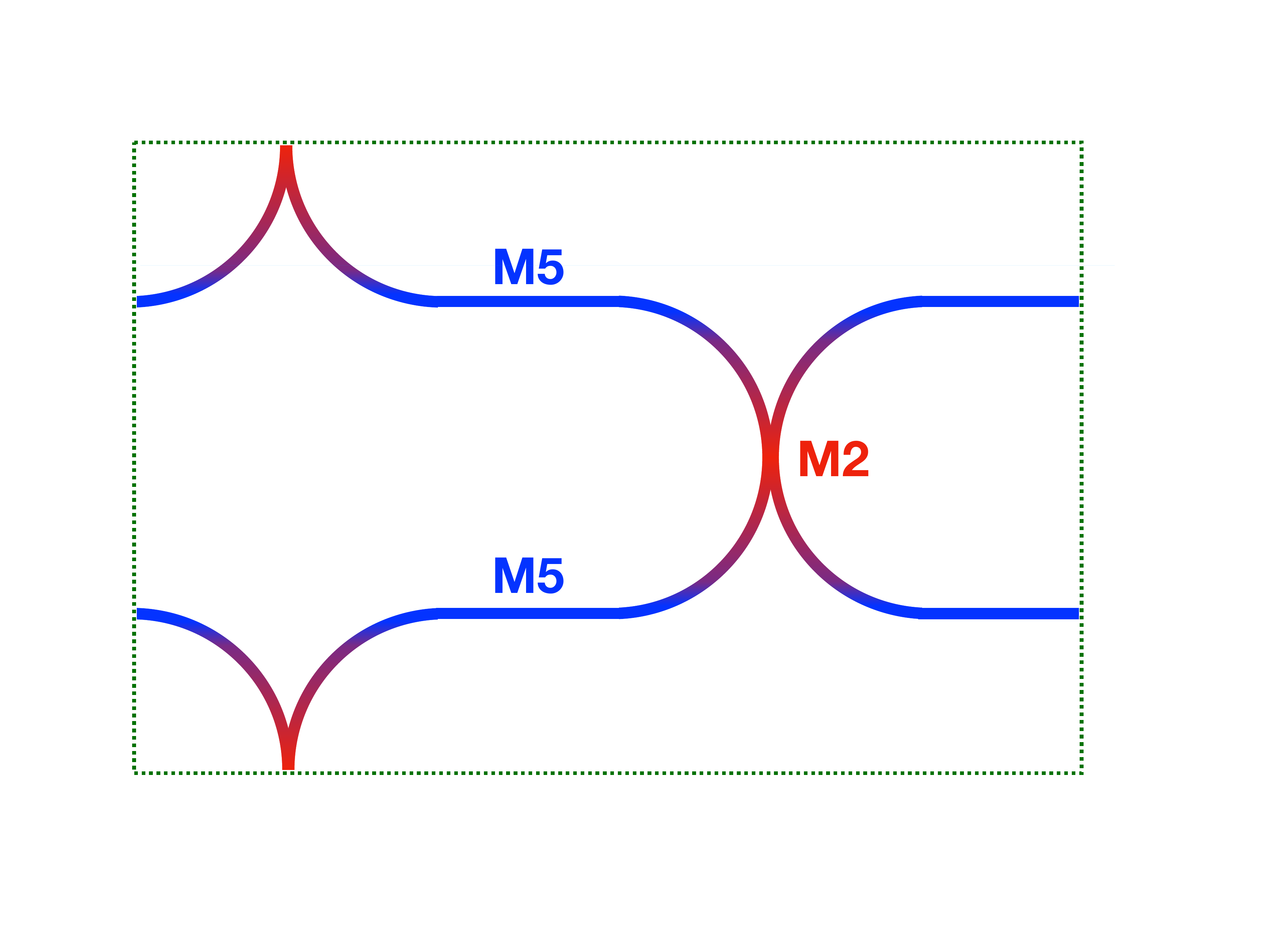}
	\caption{The fractionation of M2 branes into strips and the super-maze: Before the fractionation  (left panel) the M2 brane does not interact with the M5 branes, and can be freely taken away. After the fractionation, each strip of the M2 branes can move independently, giving the na\"ive configuration in the middle panel. However, the M2 strips pull on the M5 brane, creating the super-maze depicted in the right panel.	}
	\label{fig:flux_before_after1}
\end{figure}

The first step in our endeavor is to understand the backreaction of the M2 strips ending on a single M5 brane. We show that their behavior is similar to that of the {\it Callan-Maldacena spikes} describing backreacted F1 strings ending on D3 branes \cite{Callan:1997kz}. Since the M5 branes and the M2 branes extend along a common direction, the M2 branes will now form ``furrows'' on the M5 brane worldvolume, whose transverse section will look like a Callan-Maldacena spike. A key feature of these bound-states is that they preserve 8 supersymmetries, but if one zooms on a piece of the spike or of the furrow, one finds that locally there are 16 preserved supersymmetries, as a result of the presence of extra ``dipolar'' charges. 

Besides the infinite M5-M2 brane furrow, one can consider more complicated bound state of multiple M5 branes and multiple M2 strips stretching between them (like the system in Figure \ref{fig:M2_fractionated_M5}). This will result in a complicated maze of furrows, that connect these M5 branes. This super-maze preserves the same supersymmetries and the M5 branes and M2 branes whose charges it carries, but if one zooms in on a piece of maze, one expects that the supersymmetry will be locally enhanced from 8 to 16 supercharges.

Note that each of the $N_1 N_5$ M2 brane strips whose pull on the M5 branes gives the super-maze can be located at an arbitrary position inside the $T^4$ or $K3$ wrapped by the M5 branes. Therefore, the dimension of the moduli space of super-mazes is $4 N_1 N_5$. This matches, as expected, the dimension of the moduli space of the D1-D5 (F1-NS5) system deformations that preserve rotational invariance in the transverse space \cite{Rychkov:2005ji,CabreraPalmer:2004asc,Bena:2014qxa}

The second step in our endeavour is to add momentum to the super-maze, in order to construct a brane bound-state configuration that has 16 supercharges locally and that carries the charges of a black hole with a macroscopically-large event horizon. To do this, we will first construct the two-charge bound states formed by M2 branes and momentum, and by M5 branes and momentum. The former is the M-theory uplift of the F1-P system, whose solutions have been described in supergravity in  \cite{Dabholkar:1995nc}. The momentum is carried by the transverse oscillations of the M2 branes and, if one zooms in near a piece of the momentum-carrying M2 brane one finds that the supersymmetry is enhanced to 16 supercharges.

Similarly, the M5 branes can carry momentum by transverse fluctuations, that we can restrict to be oriented only along the M-theory direction, so that the resulting solution is spherically symmetric in the non-compact spacetime directions. This system is the uplift of the NS5-P-D0-D4 solution found in \cite{Bena:2022sge}. This solution also preserves 8 supersymmetries, but locally the supersymmetry is enhanced to 16. For both the M2-P and the M5-P system, this is ensured by the presence of dipolar charges, which can be thought of as the ``glue'' needed to construct the bound states of two-charge system. Of course, for the M5-P system one can consider other types of glue, coming for example from 2 species of M2 branes inside the M5-brane worldvolume. The resulting configuration is called a magnetube, and its supergravity solution was constructed in  \cite{Mathur:2013nja,Bena:2013ora}.

The main result of this paper is to identify the ingredients needed to construct the bound states of the NS5-F1-P Type IIA system and of its M-theory M2-M5-P uplift. These bound states describe the DVV little strings carrying momentum in the regime of parameters where the brane interactions are taken into account. We show that there exists a supersymmetry projector corresponding to a brane configuration that has 16-supersymmetries locally and 4 globally, and which describes the zooming in on a piece of the momentum-carrying M2-M5 maze. Besides the M2, M5 and P charges of the black hole, this system has 6 other dipolar charges, which are necessary to form the glue that transforms these branes into a bound state.

The entropy of the DVV little strings carrying momentum reproduces (upon taking into account all bosonic and fermionic polarizations) the entropy of the F1-NS5-P black hole \cite{Dijkgraaf:1996cv}, and our result shows that the microstates carrying this entropy correspond (upon taking brane interactions into account) to a momentum-carrying super-maze whose supersymmetry is enhanced everywhere to 16 supercharges locally. 

It is important to emphasize that the local enhancement of supersymmetry to 16 supercharges is the hallmark of the existence in certain duality frames of smooth supergravity solutions that result from the backreaction of these configurations and, more generally, of the absence of event horizons. We will explain the connection between local enhancement of supersymmetry and smooth horizonless solutions in more detail in Section \ref{section5}. Confirming that the entropy of this black hole comes from horizonless super-mazes would constitute a proof of the fuzzball proposal for three-charge supersymmetric black holes in String Theory, and we are looking forward to the construction of the fully-backreacted solution corresponding to the brane microstates we have discovered.

Our result points towards a change of strategy in the fuzzball/microstate geometry programme of constructing horizonless solutions dual to microstates of string-theory black holes. Until now, the  strategy of this programme has been to ``blow up'' the delta-function source of the harmonic functions of the branes making up the black hole, and replace it by an extended source in the \textit{non-compact} spatial dimensions. This has resulted in a huge pl\ae thora of solutions \cite{Giusto:2004kj,Bena:2005va,Berglund:2005vb,Bena:2006is,Bena:2006kb,Bena:2007kg,Bena:2007qc,Bena:2008wt,Bena:2010gg,Bena:2011dd, Bianchi:2016bgx, Bianchi:2017bxl, Heidmann:2017cxt, Bena:2017fvm, Avila:2017pwi,Tyukov:2018ypq, Bena:2015bea,Bena:2016ypk,Bena:2017geu,Bena:2017xbt,Bena:2018mpb,Ceplak:2018pws,Heidmann:2019zws,Heidmann:2019xrd,Mayerson:2020tcl, Shigemori:2020yuo, Bena:2020yii,Bena:2020iyw,Giusto:2020mup,Houppe:2020oqp,Ganchev:2021pgs, Ganchev:2021iwy}, all of which break the spherical symmetry of the black-hole horizon. However, the connection between these solutions and the microstates that give rise to the black-hole entropy at weak coupling is difficult to establish, and has only been worked out for superstrata, whose  entropy is parametrically smaller than that of the black hole \cite{Shigemori:2019orj,Mayerson:2020acj}. Furthermore, all the known superstrata have at least one unit of angular momentum in one of the non-compact angular directions in which supersymmetric black holes cannot rotate\footnote{The five-dimensional supersymmetric three-charge black holes can have finite $J_L$, but their $J_R$ must be zero. In contrast superstrata always have $J_R \neq 0$.}. 

Our work points out a new route for constructing microstate geometries that solves these two challenges at the same time: the momentum-carrying super-maze preserves the same spacetime spherical symmetry as the black-hole solution, and it is directly connected to DVV fractionated strings that give rise to the entropy of the F1-NS5-P black-hole in Type IIA String Theory. Furthermore, as we will explain in Section  \ref{section5} the fact that locally the supersymmetry is enhanced to 16 supersymmetries indicates that the fully-backreacted super-mazes will give rise to smooth horizonless black hole microstate geometries, and will not have an event horizon. 

In Section \ref{sec:two_charge} we review the construction of two-charge bound states and the role of branes that act as ``glue'' and transform singular configurations of branes into bound states preserving locally 16 supercharges. In Section 3 we describe the construction of the new three-charge bound state, which preserves 16 supercharges locally and is a piece of the super-maze coming from the backreaction of DVV black-hole microstates. In Section \ref{section4} we explain the link between the projector and the local orientation of the branes that make up the super-maze, and confirm our construction by showing that the energy of the super-maze saturates the BPS bound. In Section \ref{section5} we discuss the relation between smooth horizonless supergravity solutions and brane configurations preserving locally 16 supercharges, and argue that the backreaction of the super-maze will give rise to bubbling horizonless solutions. In Appendix 1 we collect the projectors corresponding to branes, strings, KK Monopoles and momentum in String and M Theory.

{\bf Note on nomenclature:} Throughout this paper we will refer to two- and three-charge systems as systems of two or three sets of branes that preserve 8 respectively 4 common supersymmetries and that exert no force on each other. Thus, a system of D5 branes and parallel D1 branes can be properly called a two-charge system, but a system of D3 branes and parallel D1 branes is not: the D1 branes are attracted to the D3 branes and form a bound state that has 16 supercharges everywhere and is T-dual to a single oblique stack of parallel D2 branes. Similarly a D7 brane and a parallel D1 brane do not constitute a two-charge system, because the D1 branes run away from the D7 branes.

\section{Making two-charge bound states out of strings and branes}
\label{sec:two_charge}

The vacuum of Type II String Theory preserves 32 supersymmetries. Adding excitations such as strings or branes decreases the number of preserved supersymmetries. Indeed, one can derive using the BPS equations that the presence of branes imposes a constraint on the Killing spinor $\epsilon$:
\begin{equation}
 	P \, \epsilon ~=~ -\epsilon \,,\qq{or equivalently} \Pi \, \epsilon ~\equiv~ \frac12 (1 + P) \, \epsilon ~=~ 0\,,
 	\label{eq:bps_constraint_simple}
 \end{equation} 
 where $P$ is a traceless {\em involution} ($P^2 = 1$), typically a product of gamma matrices, that depend on the exact type and orientation of the object considered. Thus $\Pi$ is a projector, verifying $\Pi^2 = \Pi$. A list of the involutions corresponding to branes, strings, solitons and momentum waves is given in Appendix \ref{sec:projectors_and_involutions_for_branes}. The constraint \eqref{eq:bps_constraint_simple} divides the number of preserved global supersymmetries by two.

If one considers configurations with several types of branes whose supersymmetries are compatible, the constraints add up. For example, for a two-charge system, the Killing spinor must respect
\begin{equation}
	\Pi_1 \, \epsilon ~=~ 0 \,,\qand \Pi_2 \, \epsilon ~=~ 0 \,.
\end{equation}
In other words, the Killing spinor must lie in the intersection of the kernels of $\Pi_1$ and $\Pi_2$. The dimension of this intersection is the number of preserved global supersymmetries (8).

The number of states of a two-charge system is however much larger than one can surmise by considering the individual motion of its component branes. Indeed, the branes can form bound states\footnote{In the D0-D4 system for example, the individual motion of the branes corresponds to the Coulomb branch, where the branes do not form a bound state. However, the large degeneracy of the system comes from the Higgs branch, which describes bound states of D0 branes inside the D4 branes.}, which contain more fields than those of the na\"ive multi-brane solution. These fields can be thought of as coming from the dipolar branes that act as the ``glue'' needed to form the bound state, and which also give rise to a local enhancement of the number of preserved supersymmetries to 16.

For a general bound state, the Killing spinor satisfies
\begin{equation}
	\hat\Pi \, \epsilon ~\equiv~ \frac12 \qty(1 + \alpha_1 P_1 + \dots + \alpha_n P_n) \, \epsilon ~=~ 0 \,,
\end{equation}
where $P_i$ are the traceless involutions associated to the branes whose charges the bound state has and where, for each species of brane, $i$, the coefficient $\alpha_i$ is the ratio between the charge density corresponding to this brane, $Q_i$,\footnote{Note that the dependence in the string coupling, $g_s$, enters in the $Q_i$'s.} and the mass density of the full bound state, $M$:
\begin{equation}
\displaystyle
\alpha_i \equiv \frac{Q_i}{ M}  \,. \label{QMratio}
\end{equation}
Hence, the projector can be written as 
\begin{equation}
	 \hat\Pi  ~\equiv~ \frac{1}{2 M} \qty(M + Q_1 P_1 + \dots + Q_n P_n) \,,
\end{equation}

The number of preserved supersymmetries is now the dimension of the kernel of $\hat\Pi$. This operator is in general not a projector, but when it is, the configuration preserves 16 global supersymmetries.

It is thus possible to reveal the extra dipole charges needed to construct the bound states of a two- or three-charge system by finding involutions corresponding to suitable branes and tuning their charges to make $\hat\Pi$ a projector. For a given bound state, the solution for $\alpha_i$ is often not unique: There often exists a whole moduli space of values of the charges that make $\hat\Pi$ a projector. One can then imagine varying these charge densities along the internal dimensions of the bound state, so that the constraint becomes:
\begin{equation}
	\hat\Pi(\vec{x}) \, \epsilon(\vec{x}) ~\equiv~ \frac12 \Big[1 + \alpha_1(\vec{x}) P_1 + \dots + \alpha_n(\vec{x}) P_n \Big] \epsilon(\vec{x}) ~=~ 0 \,,
	\label{eq:local_susy_general}
\end{equation}
where $\vec{x}$ denotes the internal dimensions of the bound state. Doing so, the number of \emph{local} preserved supersymmetries is still 16. However, the number of \emph{global} supersymmetries can be much less:  it is the dimension of the common kernel to the projectors at all possible values of $\vec{x}$. The global Killing spinor does not depend on the position $\vec{x}$, and it must satisfy
\begin{equation}
 	\forall\, \vec{x}\,, \quad \hat\Pi(\vec{x}) \, \epsilon ~=~ 0 \,,
 	\label{eq:susy_constraint_all_x}
 \end{equation} 
 where $\epsilon$ must be constant.

 A common way to ensure that at least some amount of supersymmetry is preserved globally is to rewrite the projectors as
 \begin{equation}
 	\hat\Pi(\vec{x}) ~=~ f_1(\vec{x}) \Pi_1 + \dots + f_k(\vec{x}) \Pi_k
 	\label{eq:global_susy_general}
 \end{equation}
 where $\Pi_1, \dots, \Pi_k$ are commuting projectors, and $f_1, \dots, f_k$ can be any matrix-valued functions. Then, satisfying \eqref{eq:susy_constraint_all_x} is equivalent to
 \begin{equation}
 	\Pi_1 \, \epsilon ~=~ \dots ~=~ \Pi_k \, \epsilon ~=~ 0\,,
 \end{equation}
so the number of preserved global supersymmetries is the dimension of the intersection of the kernels of $\Pi_1,\dots,\Pi_k$.

 When constructing bound states, one typically starts with the set of global charges and their projectors, $\Pi_1, \dots, \Pi_k$. Combining \eqref{eq:local_susy_general} with \eqref{eq:global_susy_general} then leads to constraints on the charges of each constituent, $Q_i$, and hence on $\alpha_i$.

The distinction between local and global supersymmetries is important, and is at the core of the results of this paper. As we explained in the Introduction, by constructing  two- and three-charge  bound states preserving 16 local supersymmetries, we ensure that we construct microstates of these two- or three-charge systems and that furthermore their backreaction will not give rise to an event horizon. 
 
This bound-state making philosophy was first used to conjecture the existence of superstrata \cite{Bena:2011uw}, but the method presented here is a generalisation of that of \cite{Bena:2011uw}, where an orthogonal momentum, P$(\psi)$, was imposed to be one of the dipoles. 
In the following Subsection, we will first illustrate the bound-state making philosophy with several examples of two-charge bound states.

\subsection{The F1-P bound state}
\label{sec:F1-P_bound_state}

Consider an F1-P system where the strings wrap a compact direction, $y$, and momentum is also along $y$ direction. The involutions and projectors associated to them are (see Appendix \ref{sec:projectors_and_involutions_for_branes}):
\begin{equation}
\begin{aligned}
	P_{\mathrm{F1}(y)}~&=~ \Gamma^{0y} \sigma_3 \,, \qquad \Pi_{\mathrm{F1}(y)}~=~\frac12 (1+ P_{\mathrm{F1}(y)}) \,,
	\\
	P_{\mathrm{P}(y)} ~&=~ \Gamma^{0y} \,, \qquad \Pi_{\mathrm{P}(y)} ~=~\frac12 (1+ P_{\mathrm{P}(y)}) \,.
\end{aligned}	
\end{equation}

When the F1 and P do not form a bound state, the constraints on the Killing spinor add up
\begin{equation}
	\Pi_{\mathrm{F1}(y)}\epsilon ~=~ \Pi_{\mathrm{P}(y)}\epsilon ~=~ 0\,.
\end{equation}
and the system preserves 8 supersymmetries everywhere.

It is possible to form a bound state possessing the same global charges as this system, but preserving locally 16 supersymmetries. In order to do so, one needs to add dipolar transverse strings and momentum which we choose to be along a single transverse direction inside the $T^4$, that we call $x_1$ (more complicated choices are also possible but not illustrative for our purpose here):  
\begin{equation}
\begin{aligned}
	P_{\rm F1(1)} ~&=~ \Gamma^{01} \sigma_3 \,, \qquad \Pi_{\rm F1(1)} ~=~\frac12 (1+ P_{\rm F1(1)}) \,,
	\\
	P_{\mathrm{P}(1)}  ~&=~ \Gamma^{01} \,, \qquad \Pi_{\mathrm{P}(1)}  ~=~\frac12 (1+ P_{\mathrm{P}(1)} ) \,.
\end{aligned}	
\end{equation}

The objective is to construct a local projector, $\Pi_{\text{\rm F1-P bound}}$, that can be written in two ways:
\begin{align}
	\Pi_{\text{\rm F1-P bound}} ~&=~  \frac12 \qty(1 + \alpha_1 P_{\mathrm{F1}(y)}+ \alpha_2 P_{\mathrm{P}(y)}+ \alpha_3 P_{\rm F1(1)}+ \alpha_4 P_{\mathrm{P}(1)}  ) \label{eq:F1P_local_form}
	\\
	&=~ f_1 \Pi_{\mathrm{F1}(y)}+ f_2 \Pi_{\mathrm{P}(y)}\label{eq:F1P_global_form}
\end{align}
where $\alpha_1,\dots,\alpha_4$ are real numbers, and $f_1,f_2$ are matrices.

The equation $\Pi_{\text{F1-P bound}}^2 = \Pi_{\text{F1-P bound}}$ first leads to
\begin{equation}
	\alpha_1^2 + \alpha_2^2 + \alpha_3^2 + \alpha_4^2 ~=~ 1 \,,\qquad \alpha_1 \alpha_4 + \alpha_2 \alpha_3 ~=~ 0 \,,
\end{equation}
while equalizing \eqref{eq:F1P_local_form} and \eqref{eq:F1P_global_form} leads to
\begin{equation}
	\begin{aligned}
		\alpha_3 + \alpha_4 ~=~ 0 \,,&\qquad \alpha_1 + \alpha_2 ~=~ 1 \,,
		\\
		f_1 ~=~ \alpha_1 - \alpha_4 \Gamma^{y1} \sigma_3 \,,&\qquad f_2 ~=~ \alpha_2 - \alpha_3 \Gamma^{y1} \sigma_3 \,. 
	\end{aligned}
\end{equation}
The solution given here for $f_1$ and $f_2$ is not unique.

Solving these equations is straightforward, the solution depends on the choice of an arbitrary angle, $\theta$:
\begin{equation}
\begin{aligned}
	\Pi_{\text{F1-P bound}} ~&=~ \frac12 \qty[1 + c^2 P_{\mathrm{F1}(y)}+ s^2 P_{\mathrm{P}(y)}+ cs P_{\rm F1(1)} - cs P_{\mathrm{P}(1)} ]
	\\
	&=~ c \qty(c + s\,\Gamma^{y1}\sigma_3) \Pi_{\mathrm{F1}(y)}~+~ s \qty(s - c\,\Gamma^{y1} \sigma_3) \Pi_{\mathrm{P}(y)} \,,
\end{aligned}
\end{equation}
where $c \equiv \cos\theta$ and $s \equiv \sin\theta$.

Geometrically, the angle $\theta$ corresponds to the inclination of the string in the $(y, x_1)$ plane. If $\theta$ is constant, the configuration is a straight string tilted in the $(y, x_1)$-plane, with transverse momentum. This transversely boosted F1 string preserves 16 supersymmetries.\footnote{For an illustration, see Figure  2 and Figure 3 in \cite{Bena:2022sge}.} In the limit $\theta = 0$, this is a pure F1 string along $y$, and when $\theta = \pi/2$ this is a pure momentum wave along $y$.

One can bend the string by allowing $\theta$ to vary along it. The resulting configuration still preserves 16 supersymmetries locally, but only 8 globally.

\subsection{The NS5-P bound state}

The same exercise can be done for the NS5-P system in type IIA. We start with NS5 branes extending along the directions $y, x_1, \dots,x_4$, and momentum along $y$. The involutions associated to them are:
\begin{equation}
	P_{{\rm NS5}(y1234)} ~=~ \Gamma^{0y1234} \,,\qquad P_{\mathrm{P}(y)} ~=~ \Gamma^{0y}\,.
\end{equation}

Once again, if these constituents  do not form a bound state the configuration preserves 8 supersymmetries. They can also form bound states that preserve locally 16 supersymmetries. Contrary to the fundamental string, the NS5-brane does not need to bend in the transverse directions to carry momentum. To make the bound state, one possibility is to use internal dipolar D4-branes (extending along the directions $x_1,\dots,x_4$) and D0-branes  \cite{Bena:2022sge}:
\begin{equation}
	P_{\rm D4(1234)} ~=~ \Gamma^{01234} i \sigma_2 \,,\qquad  P_{\rm D0} ~=~ \Gamma^0 i \sigma_2 \,.
\end{equation}
Note that this is not the only possible choice of dipoles. We can also form an F1-NS5 bound state by adding as ``glue'' two orthogonal sets of D2 branes. This system can be obtained from the one we have by two T-dualities along the NS5 internal directions that are not wrapped by the F1 strings. Its M-theory uplift is know as the magnetube \cite{Bena:2013ora, Mathur:2013nja}.

Another possibility to construct bound states with P and NS5 charges is to put a momentum-carrying transverse wave on the NS5 brane. This configuration can easily be obtained by dualizing the F1 strings with a transverse momentum wave described above and its ``glue'' consists of a dipolar NS5 charge and angular momentum. This solution breaks the spherical symmetry of the black-hole solution. Since in this paper we are interested in constructing bound states that respect this spherical symmetry and can describe locally the backreaction of the DVV microstates, we will describe in detail the brane bound states created using D0-D4 glue.

 One needs to construct a projector satisfying
\begin{align}
	\Pi_{\text{NS5-P bound}} ~&=~  \frac12 \qty(1 + \alpha_1 P_{{\rm NS5}(y1234)} + \alpha_2 P_{\mathrm{P}(y)}+ \alpha_3 P_{\rm D0}+ \alpha_4 P_{\rm D4(1234)} ) \label{eq:NS5P_local_form}
	\\
	&=~ f_1 \Pi_{{\rm NS5}(y1234)} + f_2 \Pi_{\mathrm{P}(y)} \label{eq:NS5P_global_form}
\end{align}
as well as the usual condition on projectors $\Pi_{\text{NS5-P bound}}^2 = \Pi_{\text{NS5-P bound}}$, for some real numbers $\alpha_1, \dots, \alpha_4$ and matrices $f_1,f_2$.\footnote{It is not necessary to do this computation again. One can find the result by dualizing the F1-P system (if the directions are not compact, a T-duality can be seen as a solution-generating technique rather than a proper duality). The duality chain is $T_1-S-T_{1234}-S-T_1$. }

The solution to this system is:
\begin{equation}
\begin{aligned}
	\Pi_{\text{NS5-P bound}} ~&=~ \frac12 \qty[1 + c^2 P_{{\rm NS5}(y1234)} + s^2 P_{\mathrm{P}(y)}+ cs P_{\rm D0} - cs P_{\rm D4(1234)}]
	\\
	&=~ c \qty(c + s\,\Gamma^y i \sigma_2) \Pi_{{\rm NS5}(y1234)} ~+~ s \qty(s - c\,\Gamma^y i \sigma_2) \Pi_{\mathrm{P}(y)} \,,
\end{aligned}
\end{equation}
where again $c \equiv \cos\theta$ and $s \equiv \sin\theta$.

\subsection{The NS5-F1 bound state}
\label{sec:ns5-f1-bound-state}

One can form an NS5-F1 bound state in type IIA using a similar procedure. Consider an NS5-F1 system where the NS5 extends along the directions $y, x_1,\dots,x_4$, and the string is along $y$. The involutions associated to them are
\begin{equation}
	P_{{\rm NS5}(y1234)} ~=~ \Gamma^{0y1234} \,,\qquad P_{\mathrm{F1}(y)}~=~ \Gamma^{0y} \sigma_3 \,.
\end{equation}

The bound state can be obtained from the NS5-P system by performing two T-dualities along the directions $y$ and $x_1$. Again, the choice of $x_1$ among the four torus directions is at this point arbitrary.

 The dipole charges needed to form it are D4-branes extending along the directions $y, x_2 \dots, x_4$, and D2-branes along the direction $y$ and $x_1$:
\begin{equation}
	P_{\mathrm{D4}(y234)} ~=~ \Gamma^{0y234} i \sigma_2 \,,\qquad P_{\mathrm{D2}(y1)} ~=~ \Gamma^{0y1} \sigma_ 1 \,.
\end{equation}
The projector of this bound state is
\begin{equation}
\begin{aligned}
	\Pi_{\text{NS5-F1 bound}} ~&=~ \frac12 \qty[1 + c^2 P_{{\rm NS5}(y1234)} + s^2 P_{\mathrm{F1}(y)}+ cs P_{\mathrm{D2}(y1)} + cs P_{\mathrm{D4}(y234)}]
	\\
	&=~ c \qty(c + s\,\Gamma^1 i \sigma_2) \Pi_{{\rm NS5}(y1234)} ~+~ s \qty(s - c\,\Gamma^1 i \sigma_2) \Pi_{\mathrm{F1}(y)} \,.
\end{aligned}
\label{eq:M2-M5-tilt-x1}
\end{equation}
where again $c \equiv \cos\theta$ and $s \equiv \sin\theta$, and the angle $\theta$ is a function of the coordinates $y, x_1,\dots,x_4$.

\begin{figure}[h]
\begin{center}
\includegraphics[scale=1]{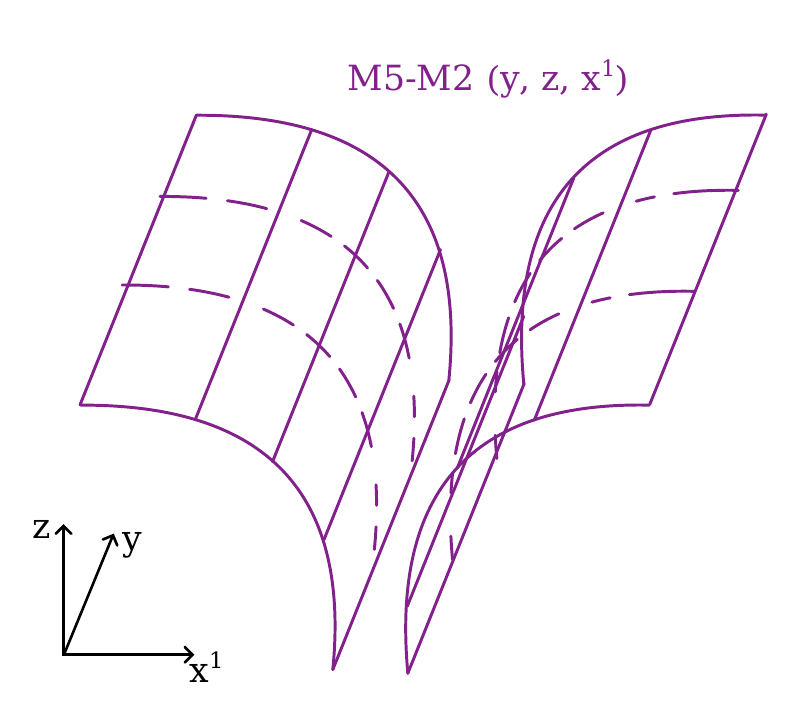}
\caption{  The backreaction of the M5-M2 bound state, projected onto the space $(y,x_1, z)$. The M2-branes pull the M5-branes, forming a furrow. The mechanism is similar to the formation of a Callan-Madacena spike in the D3-F1 brane system. }
\label{fig:M5-M2_furrow}
\end{center}
\end{figure}

The angle $\theta$ and the form of the projector have a clear geometric interpretation for the F1-P bound state (as the tilt of the string). For the NS5-F1 bound state, the interpretation is more complicated: one needs first to uplift the configuration to M-theory. The projector is then given by:
\begin{align} \label{eq:projector_M5-M2}
\Pi_{\mathrm{M5-M2}}= \frac{1}{2} \qty[ 1 
+  c^2 P_{\mathrm{M5}(y1234)} + s^2 P_{\mathrm{M2}(y\,z)} - c s P_{\mathrm{M5}(y234\,z)} + cs P_{\mathrm{M2}(y1)} ]\,.
\end{align}

 The brane system then consists of M5-branes and M2-branes sharing one common direction, $y$. The M2-branes are also extended along the M-theory circle, denoted by $z$. It is easy to see that M2 branes terminating on the M5 branes will pull them along the (previously orthogonal)  M-theory direction. This mechanism is similar to the formation of a Callan-Maldacena spike (see Fig. \ref{fig:M5-M2_furrow}). At each location on the M5-brane, the angle $\theta$ corresponds to the tilt of the brane in the $z$ direction. Of course, for a generic spike, the pull of the M2 brane will affect all the 4 directions of the NS5 brane, $(x_1,x_2,x_3,x_4) $, and the spike will be described by a complicated function of all these four variables. To obtain the bound state depicted in Figure \eqref{fig:M5-M2_furrow}, corresponding to the projector \eqref{eq:M2-M5-tilt-x1}, we can either smear the M2 branes along the directions $(x_2,x_3,x_4) $ or one can zoom in at a location of the spike where the tangent to the spike is orthogonal to $x_1$.

Another (more familiar) possibility to construct bound states with F1 and NS5 charges is to add a dipolar KKM charge (extending in the space transverse to the NS5 worldvolume and with the special direction along the F1-NS5 common direction) as well as angular momentum, J. This gives rise to a F1-NS5 supertube with KKM-J dipole charge, and its supergravity solution is the S-dual of the well-known Lunin-Mathur geometry \cite{Lunin:2001fv, Lunin:2002iz}. Much like its better known supertube cousins  \cite{Mateos:2001qs, Emparan:2001ux}, the KKM can wrap an arbitrary curve in the four dimensions transverse to the NS5 branes, and the solution preserves 8 supercharges. As reviewed in \cite{Bena:2011uw}, when one zooms near the supertube profile this configuration preserves locally 16 supercharges and this enhancement of supersymmetry comes from the presence of the KKM and angular-momentum ``glue'', and is equivalent to the fact that the supergravity solution corresponding to the F1-NS5-KKM-J supertube is smooth \cite{Bena:2008dw}. 

Starting from this two-charge bound state one can also add momentum, and build three-charge superstrata: bound states that have the same charges as an F1-NS5-P black hole and give rise to a smooth supergravity solution \cite{Bena:2015bea}. However, since our purpose in this paper is to build three-charge brane bound states that have the same charges as a black hole but that do not break the rotational symmetry of the black-hole horizon, we will not use the ``KKM-angular momentum glue'', and focus instead on the ``D4-D2 glue''. 

\subsection{The relation between the M5-M2 furrow and the Callan-Maldacena spike}
\label{sub:furrow-spike}

There are two ways to relate the M2-M5 furrow whose Type-IIA reduction gives rise to the NS5-F1 bound state to the better known F-string and D-string spikes constructed in the D3 brane worldvolume by Callan and Maldacena \cite{Callan:1997kz,Constable:1999ac}.

The first is to start with a D4-brane in the directions 1234, and an F1-string along the direction $y$, ending on the D4-brane. This picture is valid when $g_s \ll 1$. As one increases $g_s$ or the number of F1 strings, these strings pull on the D4 brane and give rise to a spike. Much like the D3-F1 spike, this D4-F1 spike can be constructed as a solution to the D4-brane DBI action. 

The M5-M2 bound state we consider is dual to the D4-F1 spike, after 11-dimensional uplift along $z$, and a flip in the coordinates $(y,z)$:
\begin{align}
	\label{eq:DualChain_spike-furrow}
	\begin{pmatrix}
		\text{D4$(x^1x^2x^3x^4)$}\\
		\text{F1$(y)$}\\
	\end{pmatrix}_{\rm IIA}\hspace*{-0.4cm}
	\xrightarrow[]{~\text{uplift on}~z~}~
	\begin{pmatrix}
		\text{M5$(z,x^1x^2x^3x^4)$}\\
		\text{M2$(z,y)$}\\
	\end{pmatrix}_{\rm M}\hspace*{-0.4cm}
	\xrightarrow{~(z,y)\text{-flip}~}~
	\begin{pmatrix}
		\text{M5$(y,x^1x^2x^3x^4)$}\\
		\text{M2$(y,z)$}\\
	\end{pmatrix}_{\rm M}\hspace*{-0.4cm} \,.
\end{align}

Another way to obtain an M2-M5 furrow -- but this time smeared over one of the internal directions -- is to construct the furrow corresponding to a D2 along the directions $0,y,z$ that ends on a D4 brane extended along $0,y,1,2,3$. From the perspective of the D4 brane DBI action, this smeared furrow has exactly the same solution as a D1-D3 spike \cite{Constable:1999ac}. This furrow can also be constructed using the non-Abelian DBI action of the D2 brane; the D2 non-commuting worldvolume fields are the same as the D6 brane fields describing a D6 brane ending on a D8 brane \cite{Bena:2016oqr}. Upon uplifting the D2-D4 furrow to 11 dimensions, one obtains a solution smeared along this direction, which is precisely the same as an M2-M5 furrow smeared along one of the M5-brane worldvolume directions:

\begin{align}
	\label{eq:DualChain_spike-furrow2}
	\begin{pmatrix}
		\text{D3$(x^1x^2x^3)$}\\
		\text{D1$(z)$}\\
	\end{pmatrix}_{\rm IIB}\hspace*{-0.4cm}
	 \xrightarrow[]{~\text{T}_y}~
	\begin{pmatrix}
		\text{D4$(y,x^1x^2x^3)$}\\
		\text{D2$(y,z)$}\\
	\end{pmatrix}_{\rm IIA}\hspace*{-0.4cm} 
	 \xrightarrow[]{~\text{uplift on}~x^4~}~
	\begin{pmatrix}
		\text{M5$(y,x^1x^2x^3\widetilde{x^4})$}\\
		\text{M2$(y,z)$}\\
	\end{pmatrix}_{\rm M}\hspace*{-0.4cm} \,.
\end{align}


\section{The three-charge NS5-F1-P bound state}
\label{sec:three_charge}

This section is devoted to the construction of the bound states of the three-charge system. As explained in the Introduction, we expect the bound state to have both the three charges of the NS5-F1-P system, but also several dipolar charges, which constitute the glue needed to construct a bound state that has locally 16 supercharges.

\subsection{Constructing the projector}

We consider the Type IIA  three-charge system with NS5-branes extending along the directions $y,x_1,\dots,x_4$, as well as F1 strings and momentum along the direction $y$. The involutions that enter in the construction of their corresponding projectors are:

\begin{equation}
	P_{{\rm NS5}(y1234)} ~=~ \Gamma^{0y1234} \,,\qquad P_{\mathrm{F1}(y)}~=~ \Gamma^{0y} \sigma_3\,,\qquad P_{\mathrm{P}(y)} ~=~ \Gamma^{0y} \,.
\end{equation}

In order to form a bound state, one needs to find the dipole charges that bind these branes into a configuration with 16 supersymmetries locally. In Section \ref{sec:two_charge} we explained how to construct two-charge bound states for the F1-P, NS5-F1 and NS5-P systems: For each system, we found several pairs of dipole charges acting as a glue between the constituents to form bound states. However, upon demanding that these bound states preserve the rotational invariance of the black-hole horizon, only a limited choice of dipole-brane glue remained. 
The intuitive first attempt at constructing the NS5-F1-P three-charge bound state is to add all the six dipole charges that enter in the construction of the rotationally-invariant two-charge bound states (summarized in Table \ref{tab:NS5-F1-P_and_glue}), and to try to construct a projector. We can easily find that this only works if the dipole charges of the F1-P bound state are oriented along the same direction, $x_1$, as the dipole charges of the NS5-F1 bound state.
\begin{table}[ht]
    \centering
\begin{tabular}{|c|c|c||c|c|c|c|c|c|}
\hline
NS5$(y1234)$ & F1$(y)$      & P$(y)$       & D4$(y234)$ & D2$(y1)$ & D4$(1234)$ & D0       & F$(1)$   & P$(1)$   \\ \hline
$\bigotimes$ & $\bigotimes$ &              & $\times$   & $\times$ &            &          &          &          \\ \hline
$\bigotimes$ &              & $\bigotimes$ &            &          & $\times$   & $\times$ &          &          \\ \hline
             & $\bigotimes$ & $\bigotimes$ &            &          &            &          & $\times$ & $\times$ \\ \hline
\end{tabular}
    \caption{Each line describes a two-charge bound state whose charges are two of the three charges of the NS5-F1-P brane systems (denoted by $\bigotimes$). Each bound state contains two more dipole charges, denoted by $\times$. We attempt to construct a three-charge bound state with NS5-F1-P and all six dipole charges.}
    \label{tab:NS5-F1-P_and_glue}
\end{table}

Constructing the projector for this bound state follows the same rules as for the two-charge systems. One needs to determine the local charges, $\alpha_i$, and the matrices, $f_j$, such that the expression:
\begin{align}
 	\Pi_{\text{NS5-F1-P bound}} ~&=~ 
 	\begin{aligned}[t] 
	\frac{1}{2} \biggl[ 1 &+ \alpha_1 P_{\mathrm{NS5}(y1234)} + \alpha_2 P_{\mathrm{F1}(y)} + \alpha_3 P_{\mathrm{P}(y)} + \alpha_4 P_{\mathrm{D4}(y234)}
	\\
	 &+\alpha_5 P_{\mathrm{D2}(y1)} +\alpha_6 P_{\mathrm{P}(1)} +\alpha_7 P_{\mathrm{F1}(1)} +\alpha_8 P_{\mathrm{D4}(1234)} +\alpha_9 P_{\mathrm{D0}} \biggr] \,.
	\end{aligned} \label{eq:NS5F1P_local_projector}
	\\
	&=~ f_1 \Pi_{\mathrm{NS5}(y1234)}+ f_2 \Pi_{\mathrm{F1}(y)}+ f_3 \Pi_{\mathrm{P}(y)}\,, \label{eq:NS5F1P_global_projector}
\end{align}
is a projector ($\Pi_{\text{NS5-F1-P bound}}^2 = \Pi_{\text{NS5-F1-P bound}}$) and moreover, as the second line illustrates, is compatible everywhere with the supersymmetries of the NS5-F1-P system.

From \eqref{eq:NS5F1P_local_projector} and \eqref{eq:NS5F1P_global_projector} we find:
\begin{equation}
	\begin{aligned}
		\alpha_1 + \alpha_2 + \alpha_3 ~=~ 1 \,, \quad & \alpha_4 - \alpha_5 ~=~ 0 \,,\quad \alpha_6 + \alpha_7 ~=~ 0 \,,\quad \alpha_8 + \alpha_9 ~=~ 0 \,,
		\\
		f_1 ~&=~ \alpha_1 ~+~ \alpha_4 \, \Gamma^1 i\sigma_2 ~-~ \alpha_8 \, \Gamma^y i \sigma_2 \,,\\
		f_2 ~&=~ \alpha_2 ~-~ \alpha_5 \, \Gamma^1 i\sigma_2 ~-~ \alpha_6 \, \Gamma^{y1} \sigma_3 \,,\\
		f_3 ~&=~ \alpha_3 ~-~ \alpha_9 \, \Gamma^y i\sigma_2 ~-~ \alpha_7 \, \Gamma^{y1} \sigma_3 \,.
	\end{aligned}
\end{equation}
Here again the values of the functions $f_j$ are not unique. The equation  $\Pi_{\text{NS5-F1-P bound}}^2 = \Pi_{\text{NS5-F1-P bound}}$ leads to:
\begin{gather}
	\sum_{i=1}^9 \alpha_i^2 ~=~ 1 \,, \\
	\alpha_2 \alpha_3 + \alpha_6 \alpha_7 ~=~ 0 \,,\qquad \alpha_3 \alpha_5 + \alpha_7 \alpha_9 ~=~ 0 \,, \qquad \alpha_2 \alpha_9 - \alpha_5 \alpha_6 ~=~ 0 \,,
	\\
	\alpha_3 \alpha_4 + \alpha_6 \alpha_8 ~=~ 0 \,, \qquad \alpha_1 \alpha_3 + \alpha_8 \alpha_9 ~=~ 0 \,, \qquad \alpha_1 \alpha_2 - \alpha_4 \alpha_5 ~=~ 0 \,,
	\\
	\alpha_1 \alpha_6 - \alpha_4 \alpha_9 ~=~ 0 \,, \qquad \alpha_1 \alpha_7 - \alpha_5 \alpha_8 ~=~ 0 \,, \qquad \alpha_2 \alpha_8 - \alpha_4 \alpha_7 ~=~ 0 \,.
\end{gather}

The solutions to this system can be expressed in terms of three real numbers $(a,b,c)$ satisfying $a^2 + b^2 + c^2 = 1$:
\begin{subequations} \label{alphas_vs_abc}
\begin{gather} 
	\alpha_1 ~=~ a^2 \,,\qquad \alpha_2~=~ b^2 \,, \qquad \alpha_3 ~=~ c^2 \,,
	\\
	 \alpha_4  ~=~ ab \,, \qquad  \alpha_5~=~ ab \,, \qquad  \alpha_6 ~=~ bc \,,
	\\
	\alpha_7~=~ -bc  \,, \qquad \alpha_8~=~ -ac  \,, \qquad \alpha_9~=~ ac  \,.
\end{gather}
\end{subequations}

Then the projector is:
\begin{equation}
	\Pi_{\text{NS5-F1-P bound}} ~= 
	\begin{aligned}[t]
		\frac{1}{2} \biggl[1 &+ a^2 P_{\mathrm{NS5}(y1234)} + b^2 P_{\mathrm{F1}(y)} + c^2 P_{\mathrm{P}(y)} + ab \left( P_{\mathrm{D4}(y234)} + P_{\mathrm{D2}(y1)} \right) \\
		& + bc \left( P_{\mathrm{P}(1)} - P_{\mathrm{F1}(1)} \right)
		- ac \left( P_{\mathrm{D4}(1234)} - P_{\mathrm{D0}} \right) \biggr] \,.
	\end{aligned}
	\label{eq:projector_NS5-F1-P_bound_final}
\end{equation}

\begin{figure}[h]
\begin{center}
\includegraphics[width=0.9\linewidth]{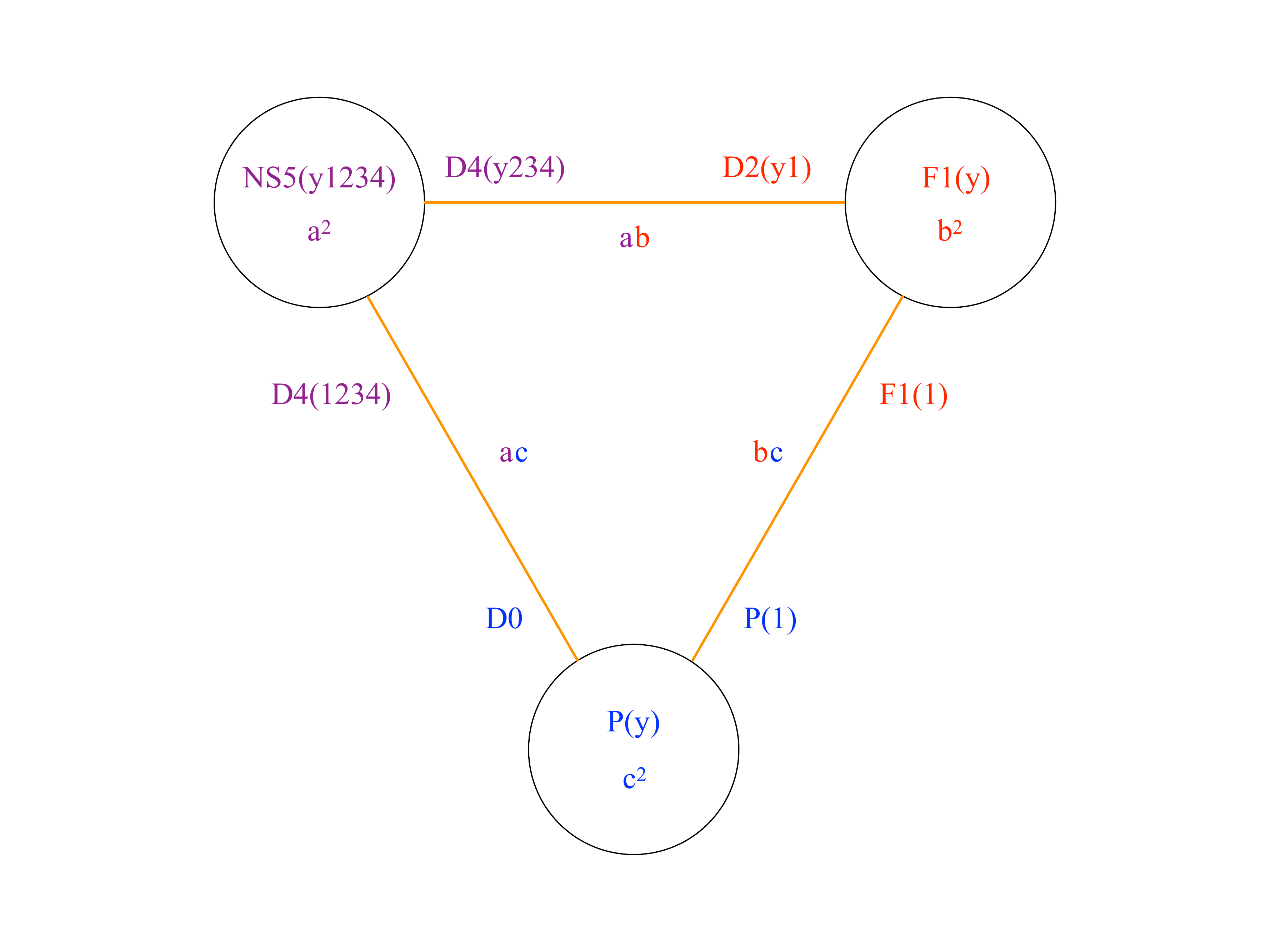}
\caption{Schematic representation of the three-charge NS5-F1-P bound state. The nodes represent the three charges of the bound state. Every combination of two nodes and the orange line joining them corresponds to a two-charge bound state, and the dipole charges and their coefficients in the projector are indicated next to the line. }
\label{fig:IIA_Trinity}
\end{center}
\end{figure}

This projector preserves locally 16 supersymmetries. We now allows the parameters $a,b,c$ to be functions of the coordinates $y,x_1,\dots,x_4, z$. The supersymmetries rotate, but the projector still preserves the 4 global supercharges of the NS5-F1-P brane system:
\begin{equation}
	\Pi_{\text{NS5-F1-P bound}} ~= 
	\begin{aligned}[t]
	& a\,(a + b \, \Gamma^1 i\sigma_2 + c \, \Gamma^y i \sigma_2) ~ \Pi_{\mathrm{NS5}(y1234)} \\
	&+ \, b\,(b - a \, \Gamma^1 i\sigma_2 - c \, \Gamma^{y1} \sigma_3) ~ \Pi_{\mathrm{F1}(y)} \\&+ \, c\,(c - a \, \Gamma^y i\sigma_2 + b \, \Gamma^{y1} \sigma_3) ~ \Pi_{\mathrm{P}(y)}\,.
 	\end{aligned}\label{eq:projector_NS5-F1-P_bound_final2}
\end{equation}
We represent the relative densities of the branes whose charges enter in this projector in Figure \ref{fig:IIA_Trinity}.

Of course, in order for the projector \eqref{eq:projector_NS5-F1-P_bound_final} to correspond to a physical brane configuration the densities of branes wrapping a certain direction should not be functions of this direction. Since these densities are related to the coefficients in the projector via equation \eqref{QMratio} this puts certain constraints on the parameters $a, b$ and $c$. These constraints will be further explained in Section \ref{section4}.

\subsection{The M5-M2-P triality}

It is also possible to uplift the three-charge bound state to M-theory, and argue that it is related to the local structure of DVV black-hole microstates.

The charges and dipole charges of the bound state have a clear M-theory origin. For simplicity we can rename the M-theory direction $x_{11}\equiv z$. As we explained in Section \ref{sec:ns5-f1-bound-state}, the two-charge bound state of F1 strings and NS5 branes can be interpreted in M-theory as the near-brane limit of the furrow created by the backreaction of M2 branes that end on M5 branes. From the perspective of the M5 brane worldvolume theory, this furrow can be constructed similarly to the Callan-Maldacena spike describing the F1 strings terminating on D3 branes. 

\begin{figure}[h]
\begin{center}
\includegraphics[width=0.9\linewidth]{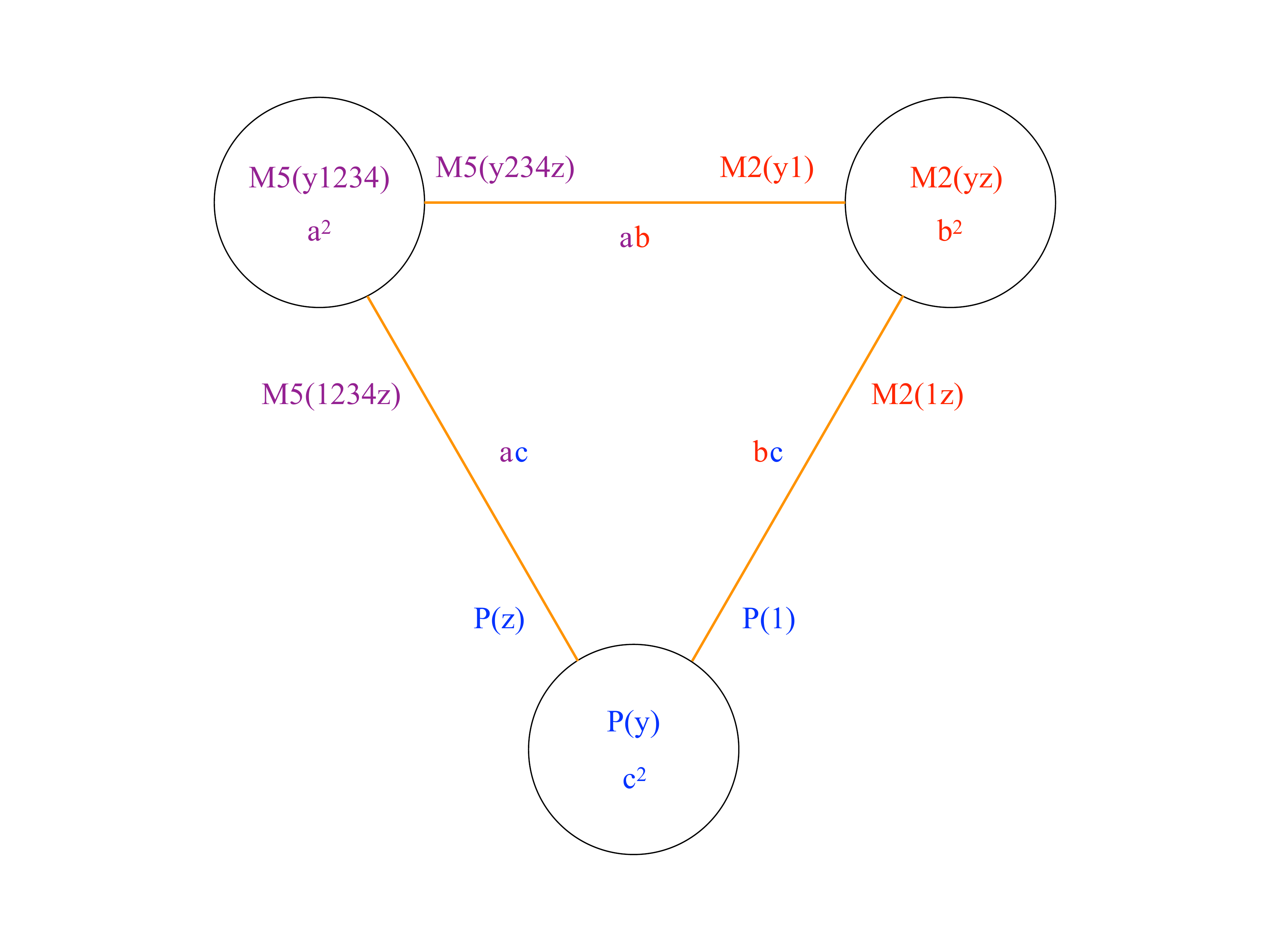}
\caption{The M-theory uplift of the NS5-F1-P bound state. The nodes represent the three charges of the bound state. Every combination of two nodes and the orange line joining them corresponds to a two-charge bound state, and the dipole charges and their coefficients in the projector are indicated next to the line.}
\label{fig:Mtheory_Trinity}
\end{center}
\end{figure}

From the M-theory perspective, the dipole branes which form the glue of the M2-M5-P bound state are also M2 and M5 branes and momentum, oriented differently. 
The NS5-brane along $(y1234)$ becomes an M5 brane along the same directions, and the F1 string along $(y)$ becomes an M2 brane along $(y,z)$. The gluing dipole branes correspond to M5 branes along $(1234z)$ and along $(y234z)$, M2 branes along $(y1)$ and along $(1z)$, and momentum along $1$ and along $z$. Figure \ref{fig:Mtheory_Trinity} reveals this triality. 

In terms of M-theory ingredients, the projector is written as:
\begin{align} \label{eq:projector_M5-M2-P}
\Pi_{\mathrm{NS5-F1-P}}= \frac{1}{2} \biggl[ 1 
+  a \hat P_{\mathrm{M5}}
+  b \hat P_{\mathrm{M2}}
+  c \hat P_{\mathrm{P}}  \biggr] \,,
\end{align}
where
\begin{align} 
\hat P_{\mathrm{M5}} &\equiv a P_{\mathrm{M5}(y1234)} - b P_{\mathrm{M5}(y234\,z)} + c P_{\mathrm{M5}(1234\,z)} \label{eq:projector_for_M5} \,,\\
\hat P_{\mathrm{M2}} &\equiv  a P_{\mathrm{M2}(y1)} + b P_{\mathrm{M2}(y\,z)} - c P_{\mathrm{M2}(1\,z)} \label{eq:projector_for_M2} \,,\\
\hat P_{\mathrm{P}} &\equiv - a P_{\mathrm{P}(z)} + b P_{\mathrm{P}(1)} + c P_{\mathrm{P}(y)} \,, \label{eq:projector_for_P}
\end{align}
and the brane involutions are all of the following form:
\begin{align} 
 & P_{\mathrm{M5}(y1234)} = \Gamma^{0y1234}  \label{eq:realprojector_for_M5}\\
 & P_{\mathrm{M2}(y1)} = \Gamma^{0y1}  \label{eq:realprojector_for_M2}\\
 & P_{\mathrm{P}(z)} =\Gamma^{0\,z}  \,. \label{eq:realprojector_for_P}
\end{align}

\section{The Brane Content of the Super-Maze}
\label{section4}

Equations (\ref{eq:projector_for_M5}), (\ref{eq:projector_for_M2}) and (\ref{eq:projector_for_P}), together with Figure \ref{fig:Mtheory_Trinity}, reveal to us the microscopic physics of the M-theory super-maze.
As we explained in Section \ref{sec:ns5-f1-bound-state}, to understand the local physics of the super-maze surface it is best to work in the $(y, 1, z)$ space, in which both the M5 branes and the M2 branes wrap nontrivial two-surfaces, and 1 denotes a torus direction orthogonal to the original M2 brane. One can see for example that equation (\ref{eq:projector_for_M5}) implies that at every location along the super-maze, the local M5 charge density in the $(y,1)$ direction (proportional to the projection of the surface of the super-maze along the $(y,1)$-plane) is equal to $a$ times the mass density of the full configuration. {We recall that the parameters $a,b,c$ have been promoted to functions of the position on the brane bound state.} One can also see that equation (\ref{eq:projector_for_M2}) implies that the {local} M2 charge in the direction $(y,z)$ (again proportional to the projection of the M2 charge of the super-maze in the $(y,z)$-plane) is equal to $b$ times the mass density.\footnote{Strictly speaking, the value is $\pm b$ , where the choice of $\pm$ depends on which one of the two unit vectors orthogonal to the M5-brane surface we choose.}

We can span the $(y, 1, z)$ space using orthonormal vectors $(u_{y}, u_1,u_z)$.
 Let $u^\perp_{\mathrm{M5}}$ be the unit vector orthogonal to the two-dimensional M5-brane surface in the $(y, 1, z)$ space. Let $u^\perp_{\mathrm{M2}}$ be its equivalent for the M2-brane, and $u_{\mathrm{P}}$ the unit vector along the direction of the momentum P.
Then, by choosing the orientation signs appropriately, one can show  that the equations (\ref{eq:projector_for_M5}), (\ref{eq:projector_for_M2}) and (\ref{eq:projector_for_P}) imply successively
\begin{align}
a &= u^\perp_{\mathrm{M5}} \cdot u_{z} \,, \qquad b = u^\perp_{\mathrm{M5}} \cdot u_{1} \,, \qquad c = u^\perp_{\mathrm{M5}} \cdot u_{y} \,,\\
a &= u^\perp_{\mathrm{M2}} \cdot u_{z} \,, \qquad b = u^\perp_{\mathrm{M2}} \cdot u_{1} \,, \qquad c = u^\perp_{\mathrm{M2}} \cdot u_{y} \,,\\
a &= u_{\mathrm{P}} \cdot u_{z} \,,\, \, \qquad \, b = u_{\mathrm{P}} \cdot u_{1} \,,\,\, \qquad \, c = u_{\mathrm{P}} \cdot u_{y} \,.
\end{align}

Hence, these equations simply imply that:
\be
u^\perp_{\mathrm{M5}} = u^\perp_{\mathrm{M2}} =  u_{\mathrm{P}} \,.
\ee

Thus, even though the super-maze has several M5 and M2 local charges pointing in different directions, when one zooms in on any particular location one finds a tilted M5 brane with parallel M2 charge dissolved in it and orthogonal momentum, which is a configuration preserving 16 supercharges. Of course, these 16 supercharges vary as one moves to a different location of the super-maze, and only 4 of them remain unchanged - the supercharges corresponding to the F1-NS5-P system whose microstates we are constructing.

Our projector also makes it clear how the energy density of the super-maze is distributed among its constituents. Before adding momentum ($c=0$), we have a static $y$-independent maze, that contains  M5 and M2 branes wrapping $y$. If one concentrates on a single furrow in the maze, the surface can de described by an equation $z = f(x_1)$. One can then parametrize $a$ and $b$ by an angle, $\beta$, that depends on $x_1$:
\begin{equation}
	a ~=~ \cos\beta \,,\qquad b ~=~ \sin\beta \,.
\end{equation}
This angle $\beta$ corresponds locally to the bending of the surface of the momentum-less maze in the $(y,z)$ plane: $\tan\beta = f'(x_1)$.

We can now compute the energy density of the momentum-less maze from its M5- and M2-brane constituent charges. Using \eqref{QMratio}, one finds 
\begin{align}
	Q^{M5}_{(y1234)} ~&=~  M \cos^2\beta \,, & Q^{M5}_{(y234z)} ~&=~ - M \cos\beta \sin\beta \,,
	\\
	Q^{M2}_{(y1)} ~&=~  M \cos\beta\sin\beta \,, & Q^{M2}_{(yz)} ~&=~  M \sin^2\beta \,,
\end{align}
where $M$ is the mass density. As usual, the square of the energy density is equal to the sum of the squares of all the charges
\begin{equation}
	M^2 ~=~ \sum_I (Q_{I})^2 \,.
\end{equation}
However, since the ratio of the M5 and M2 charges is the same as the angle of the furrow, the mass simplifies to the usual BPS mass of a two-charge system:
\begin{equation}
	M ~=~ Q^{M5}_{(y1234)} + Q^{M2}_{(yz)} \,.
\end{equation}

\begin{figure}[h]
\begin{center}
\includegraphics[scale=1]{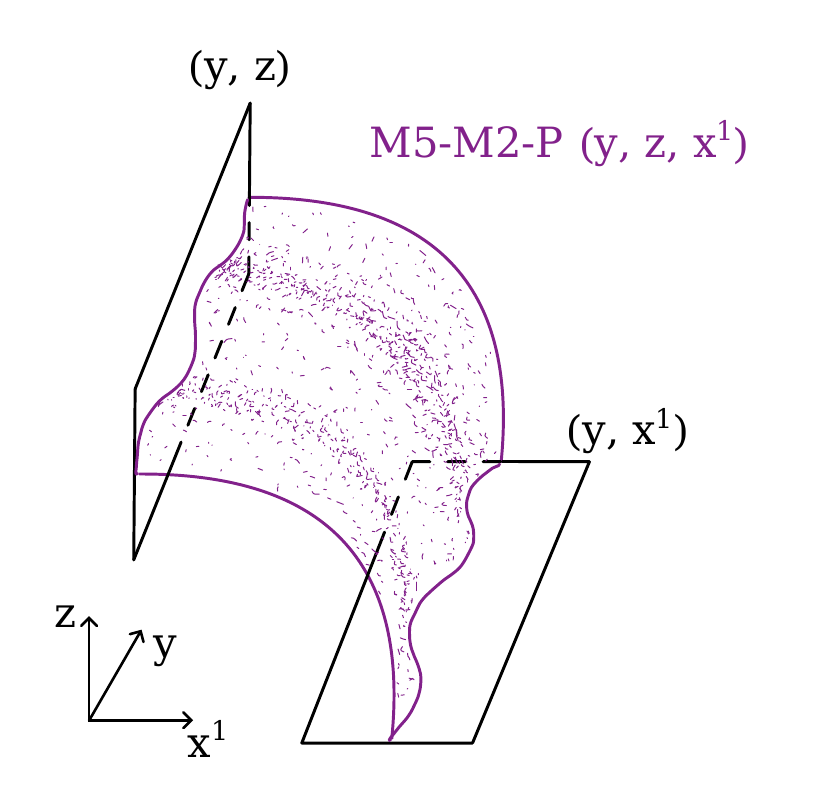}
\caption{Wiggling half-furrow.}
\label{fig:M5-M2-P_wiggle}
\end{center}
\end{figure}

If one now adds momentum, the super-maze oscillates along $y$. The furrow can now be described by a generic function of two variables
(see Figure \ref{fig:M5-M2-P_wiggle}.), and  the bending angle, $\beta$, can also become $y$-dependent.

Moreover. we also need to introduce an additional ``wiggling'' angle, $\alpha$, corresponding to the slope of the furrow waves carrying momentum along the $y$ direction. This angle can also depend on both $y$ and $x_1$. The parameters $a, b$ and $c$ can now be expressed in terms of these angles as
\begin{equation}
	a= \cos\alpha \cos\beta =  c_{\alpha} c_\beta \,,\qquad b = \cos\alpha \sin\beta = c_\alpha s_\beta \,, \qquad c = \sin \alpha = s_\alpha\,.
\end{equation}

The energy density of the momentum-carrying furrow is distributed between the branes and the momentum:
\begin{align}
	Q^{M5}_{(y1234)} ~&=~  M c_\alpha^2 c_\beta^2 \,, & Q^{M5}_{(y234z)} ~&=~ - M c_\alpha^2c_\beta s_\beta \,, & Q^{M5}_{(1234z)} ~&=~  M c_\alpha s_\alpha c_\beta \,, \label{M5chargemass}
	\\
	Q^{M2}_{(y1)} ~&=~  M c_\alpha^2 c_\beta s_\beta \,, & Q^{M2}_{(yz)} ~&=~  M c_\alpha^2 s_\beta^2 \,, & Q^{M2}_{(1z)} ~&=~ - M c_\alpha s_\alpha s_\beta \,,
	\\
	Q^{P}_{(1)} ~&=~ - M c_\alpha s_\alpha  s_\beta \,, & Q^{P}_{(z)} ~&=~  M c_\alpha s_\alpha c_\beta \,, & Q^{P}_{(y)} ~&=~ M  s_\alpha^2 \,,
\end{align}
and once again, this leads to the BPS condition for the three-charge system
\begin{equation}
	M ~=~ Q^{M5}_{(y1234)} + Q^{M2}_{(yz)} + Q^{P}_{y}\,.
\end{equation}

Note that as one moves in the $(y,1)$ plane, the projection of the M5 charge in this plane remains constant. Indeed, the original five-brane wraps the $y,1$ plane, and its charge density cannot therefore depend on $y$ or $x_1$. This appears to be in conflict with equation \eqref{M5chargemass}, but we have to realize that $M$ is the mass density of the furrow in the $(y,1)$ plane, which changes as one moves along the furrow. Hence, $Q^{M5}_{(y1234)}$ is independent on $\alpha$ and $\beta$, but $M$ depends on them:
\be
M (y,x_1)  ~=~  {Q^{M5}_{(y1234)}\over  \cos^2\alpha(y, x_1) \cos^2\beta(y, x_1)}\,.
\ee

\section{In lieu of a Conclusion: \\ Some thoughts on Super-Maze backreaction}
\label{section5}

One key question about the super-maze we discovered is whether it gives rise to a horizonless and possibly smooth solution in the regime of parameters where the classical black-hole solution exists. Na\"ively one may argue that, since the super-maze contains $N_1 N_5$ M2 brane strips crammed into a very small torus, its backreaction will give rise to a solution whose curvature is too large to be reliably described by supergravity. However, this intuition fails to take into account the fact that when branes backreact they can blow up the size of the transverse spacetime.

The key feature of the super-maze that makes us confident that its backreaction will be smooth and horizonless is the local enhancement of the supersymmetry to 16 supercharges. This is the smoking gun of the construction of the brane bound states that account for the entropy of the two-charge system.
This is perhaps best known from the physics of supertubes \cite{Mateos:2001qs,Emparan:2001ux}: A supertube can have arbitrary shape and, if one zooms in at a certain  location along this shape, one finds a brane system that preserves 16 supersymmetries. Moreover, as one moves along the supertube these supersymmetries rotate, and only a subset of 8 of them is preserved by the full configuration. When the supertube is dualized to the D1-D5 (or F1-NS5) duality frame and its two charges correspond to D1 and D5 (or F1 and NS5) branes \cite{Lunin:2001fv,Lunin:2002iz}, the presence of 16 supercharges locally is equivalent to the existence of a smooth horizonless supergravity solution \cite{Bena:2008dw}. Another examples of a two-charge brane bound states is the F1 string carrying longitudinal momentum \cite{Dabholkar:1995nc}, reviewed in Section \ref{sec:two_charge}. This solution has again 8 supercharges, but if one zooms in near the location of the momentum-carrying string one finds a solution with 16 supercharges. These supercharges rotate as one moves along the string profile, and only 8 of them remain invariant and are preserved by the whole configuration. A similar bound state can be made from any brane carrying longitudinal momentum.  

A third, slightly less known illustration of a two-charge bound state that has 16 supercharges locally is the magnetube, which has again two charges, corresponding to an M5 brane and longitudinal momentum, which are bound together by the presence of M2 brane dipole charges \cite{Bena:2013ora, Mathur:2013nja}. Finally, a fourth illustration of this phenomenon is the NS5-P bound state recently constructed in \cite{Bena:2022sge}, where the supersymmetry is enhanced locally to 16 supercharges because of the presence of dipolar D0 and D4 charges on the NS5 worldvolume. 

There also exist  brane configurations that have the same charges as those of a three-charge black hole, and again have 16 supercharges locally and only 4 globally. When the brane configurations correspond to multi-center solutions \cite{Bates:2003vx} whose centers are fluxed D6 branes (which preserve locally 16 supercharges), the solutions uplift \cite{Balasubramanian:2006gi} to the smooth horizonless bubbling solutions in eleven dimensions constructed in \cite{Bena:2005va,Berglund:2005vb}. Another three-charge brane configuration that has locally 16 supercharges is the superstratum conjectured in \cite{Bena:2011uw}, which served as inspiration for the building of superstratum supergravity solutions \cite{Bena:2015bea}.

Note that in all these systems, in the absence of the dipolar branes providing the ``glue'' and in the absence of the local enhancement of the supersymmetry to 16 supercharges, one obtains singular solutions or solutions with a horizon, which do not describe microscopic degrees of freedom of these systems, but rather ensemble averages. Thus the local enhancement of the supersymmetry is the key indication that the backreaction of the brane bound state will result in a horizonless solution that describes a pure state of the system. 

It is important to remark that the local enhancement of supersymmetry and the absence of a horizon are duality-frame-invariant phenomena. Of course, in some duality frames a smooth solution can become a singular solution, but a solution with an event horizon can never be dualized to a solution without one \cite{Horowitz:1993wt} and viceversa. 

\begin{figure}[h]
	\centering
	\includegraphics[height=100px]{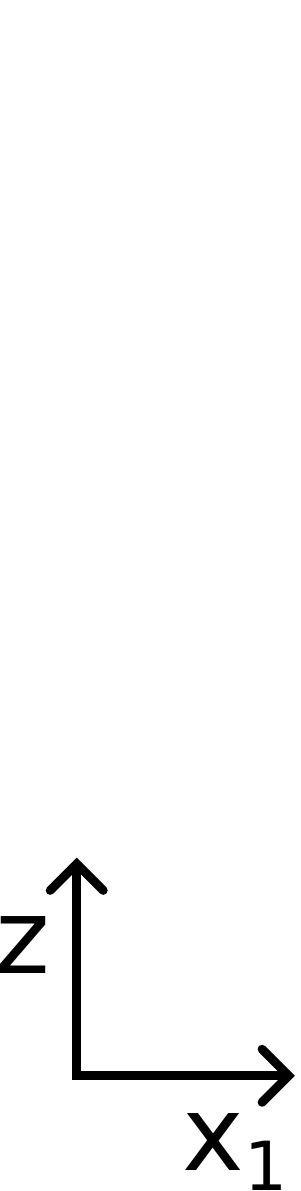}
	\includegraphics[height=125px]{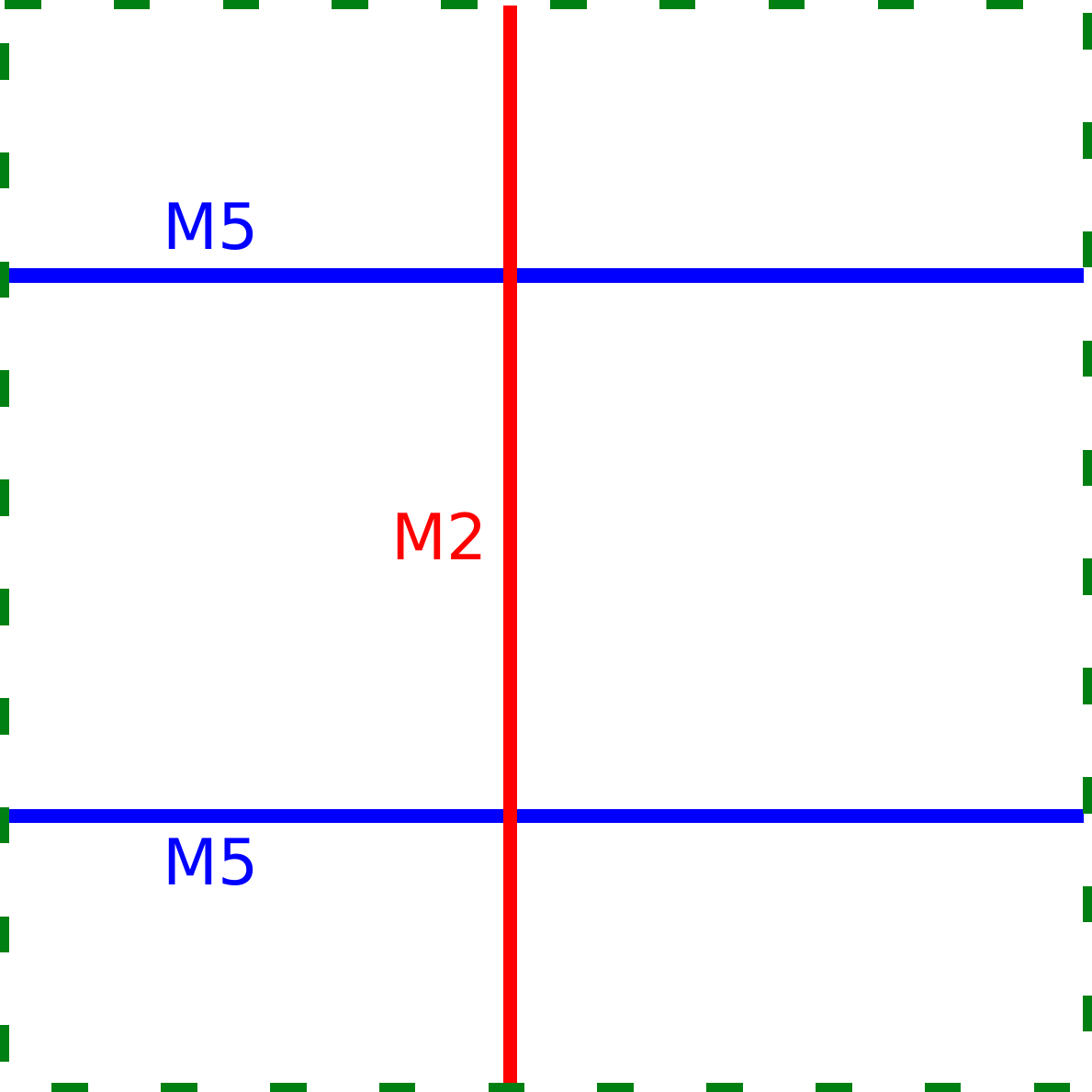}
	\hspace{.8em}
	\includegraphics[height=125px]{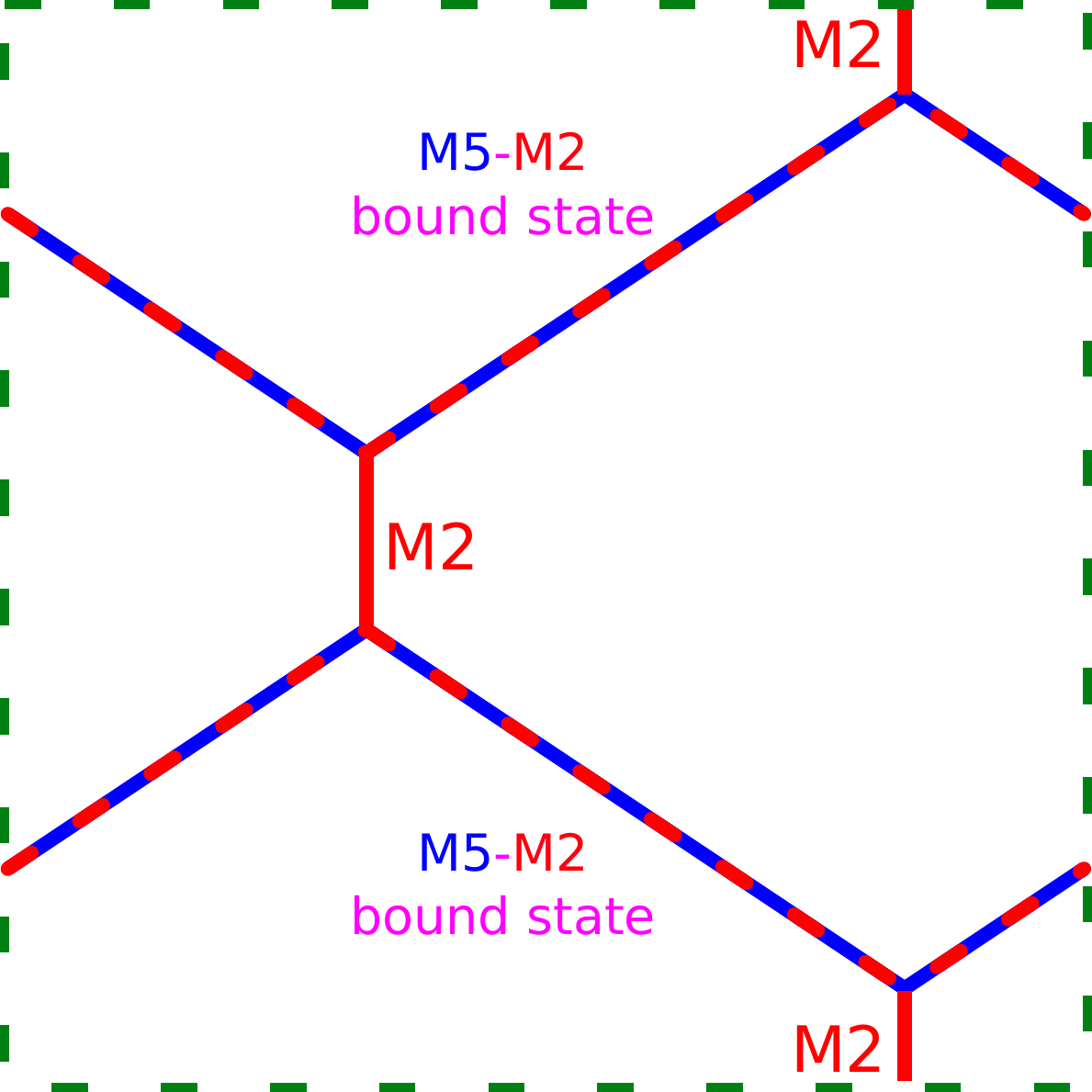}
	\hspace{.8em}
	\includegraphics[height=125px]{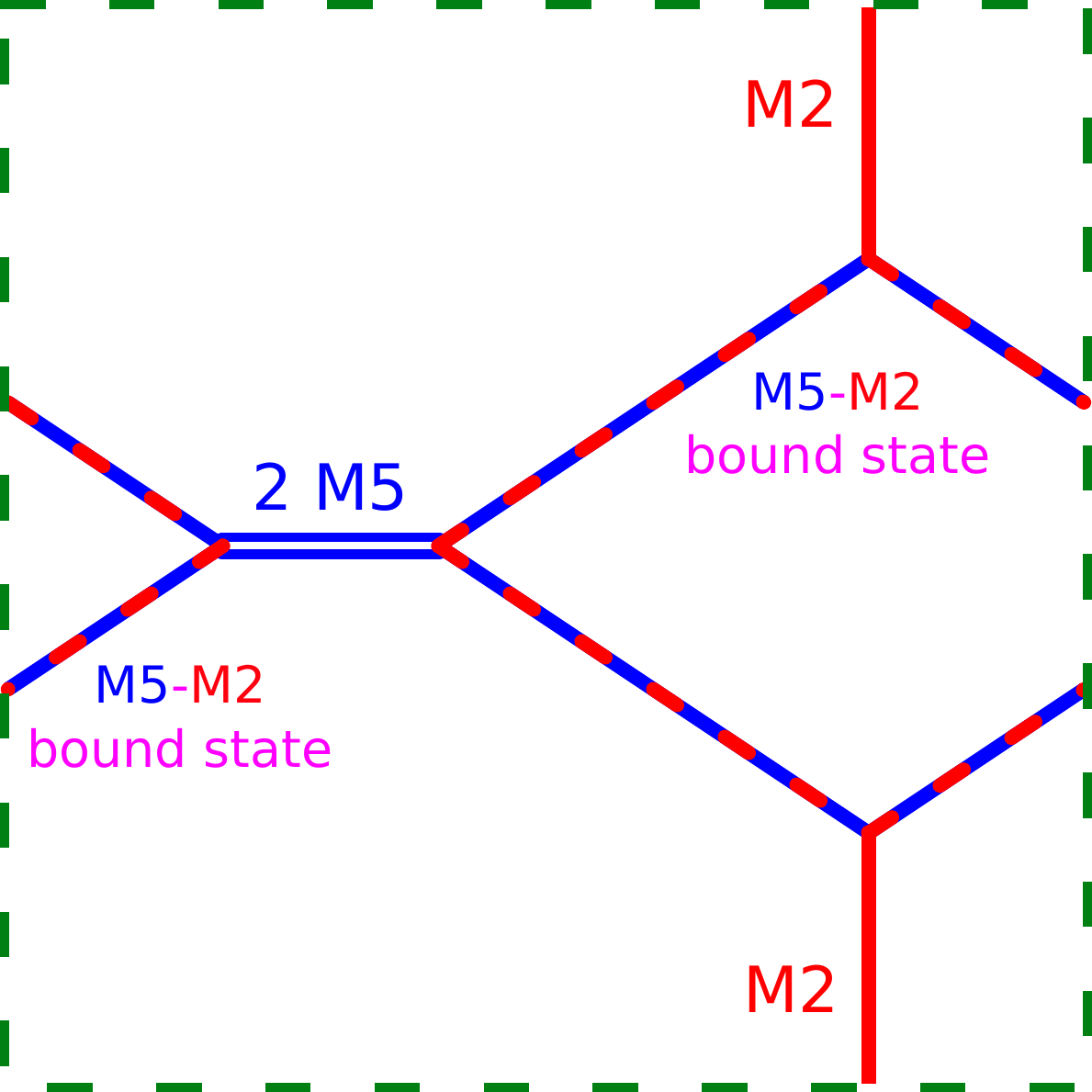}
	\caption{A super-maze made of 2 M5 branes and a single M2 brane which is smeared along three of the M5 brane worldvolume directions. Before the fractionation the M2 brane does not pull on the M5 branes, and can be freely taken away. After the fractionation (middle panel), each strip of the M2 branes deforms the M5 brane in its vicinity. As the branes move, the web depicted in the middle panel can also transform in the web depicted in the right panel, which has regions of coincident un-fluxed M5 branes. }
	\label{fig:flux_before_after}
\end{figure}

One can also speculate on how the supergravity solution corresponding to a super-maze may look. In Figure \ref{fig:flux_before_after} we illustrate the shape of a super-maze corresponding to two M5 branes extending along $x_1$ and a single fractionated M2 brane extended along $z$ (the M-Theory direction) and smeared over three of the M5-brane worldvolume directions, $x_{234}$. Before the fractionation, the M2 brane does not pull on the M5 branes; this is depicted in the left panel. Once the M2 brane gets fractionated, its components start pulling on the M5 brane. However, since the M2 brane strips have been smeared along $x_{234}$, they end on a codimension-one surface inside the M5 branes. Therefore, the pull of a fractionated M2 brane does not give rise to a spike, but rather to a wedge.\footnote{Remember that the Callan-Maldacena spike corresponds to a string ending on a codimension-three defect inside the D3 brane, and the profile of the pulled D3 brane is similar to the harmonic function in three dimensions, $1/r$. Here, the M2 branes end on a codimension-one defect, so the profile of the pulled M5 brane has the shape of the harmonic function in one dimension, $|x_1|$, and looks like a wedge.}

As one can see from the middle panel of Figure \ref{fig:flux_before_after}, when the distance between the two M5 branes is large, the configuration consists of several M5 branes wedges with dissolved M2 charge, pulled by M2 branes extended along $z$. However, the bent M5 branes can move freely along the $z$ direction, and when two opposite M5 wedges become close they  can transform into the brane web depicted in the right panel, which contains also un-fluxed coincident M5 branes. In general, a more complicated super-maze smeared over three of the M5-brane worldvolume directions will correspond to a brane web in the $(x_1, z)$-plane which has the all the three ingredients of the web in the right panel of Figure \ref{fig:flux_before_after}.

If the M2 branes are not smeared, the resulting maze does not have any ``bare'' M2 lines, but will be everywhere a fluxed M5 brane.\footnote{Even when the M2 branes are smeared, one can argue that because the distance between the M5 branes on the M-theory circle is small, the maze components will be mostly fluxed and unfluxed M5 branes.} One can then ask how the supergravity solution corresponding to this M5 super-maze will look. First, the M5 branes source a magnetic four-form whose flux on a four-sphere is constant. When the M5 branes backreact, there will be a geometric transition: this four-sphere becomes large and topologically nontrivial, while the nontrivial maze surface wrapped by the M5 branes will shrink to zero size. Thus, the maze of M5 branes will transform into a maze of bubbles with fluxes.

As we have discussed in the Introduction, the existence of super-mazes and the possibility that their supergravity solution might be smooth, represents a paradigm shift for the microstate geometry programme and for the fuzzball conjecture in general. The starting point of this conjecture is the idea that collapsing matter do not form horizons in nature, but rather transition into horizonless ``fuzzball" solutions of string theory. Standard black holes are then seen as average descriptions of the space of microstates that the stringy fuzzball matter can reach. One also expects on general grounds that some of these fuzzball solutions will have a classical limit, and will be describable purely using low-energy supergravity, as microstate geometries.

Despite the extraordinary success of the microstate geometry programme, the entropy of the solutions constructed so far, of order $\sqrt{N_1 N_5} N_P^{1/4}$ is parametrically smaller than the entropy of the three-charge black hole, $\sqrt{N_1 N_5 N_P}$. Furthermore, all the solutions that have been constructed break the spacetime spherical symmetry of the black-hole horizon, while we expect $\sqrt{5/6}$ of the black hole entropy to come from configurations that do not break this symmetry \cite{Bena:2014qxa}.\footnote{For other extremal black holes there are arguments that most of the entropy comes from such microstates \cite{Lin:2022rzw}.}

The super-maze promises to solve both these problems at the same time. On one hand, we have constructed the super-maze using the types of ``glue'' that preserve the rotational invariance of the black hole. Furthermore, the DVV microstates that we have argued to backreact into super-maze configuration correspond to momentum carriers that are purely bosonic. Hence, we expect the super-maze and its corresponding  supergravity solutions to have an entropy of order $2 \pi \sqrt{{4 \over 6} N_1 N_5 N_P}$. Furthermore, since two of the fermionic zero modes also preserve the rotational symmetry of the black-hole horizon, and the super-maze is the most general brane bound state with black-hole charges that preserves this symmetry, it is possible that the super-maze could even have an entropy of order $2 \pi \sqrt{{5 \over 6} N_1 N_5 N_P}$.

Our construction also allows us to speculate how we may try to capture the remaining part of the black-hole entropy, which comes from fermion momentum carriers that break the rotational symmetry of the black-hole horizon \cite{Bena:2014qxa}: Instead of using the super-maze glue, we could could try to use the other types of glue, and construct generalizations of the super-maze that break this rotational symmetry. 

It would be very interesting to construct the fully backreacted super-maze solutions, and to understand how this entropy is realized in supergravity. It would be also interesting to apply the ``making bound states with glue'' philosophy we used in this paper to reveal the microscopic structure of black holes in other duality frames, where microstate counting has not been done.

\noindent{\bf Acknowledgments:} We would like to thank Nejc Ceplak and Nick Warner for useful discussions. This work was supported in part by the ERC Grants 787320 ``QBH Structure'' and  772408 ``Stringlandscape.'' The work of Y.L. is also supported by the German Research Foundation through a German-Israeli Project Cooperation (DIP) grant ``Holography and the Swampland.''

\appendix

\section{Projectors and involutions for branes}
\label{sec:projectors_and_involutions_for_branes}

In this Appendix we list the involutions associated to common brane type. In Type II string theory, they are:%
\begin{align}
	P_{\mathrm{ P  }} = \Gamma^{01}\,,&\qquad
	P_{\mathrm{ F1 }} = \Gamma^{01}\sigma_3 \,, \qquad
	\notag\\
	P_{\mathrm{ NS5}}^{\mathrm{ IIA}} = \Gamma^{012345} \,,& \qquad
	P_{\mathrm{ NS5}}^{\mathrm{ IIB}} = \Gamma^{012345} \sigma_3 \,,
	\notag\\		
	P_{\mathrm{ KKM(12345;6)}}^{\mathrm{ IIA}} = \Gamma^{012345} \sigma_3 \,, &\qquad
	P_{\mathrm{ KKM(12345;6)}}^{\mathrm{ IIB}} = \Gamma^{012345} \,,
	\\
	P_{\mathrm{ D0 }} = \Gamma^0 i\sigma_2 \,, \qquad
	P_{\mathrm{ D2 }} = \Gamma^{012} \sigma_1 \,,&\qquad 
	P_{\mathrm{ D4 }} = \Gamma^{01234} i\sigma_2 \,, \qquad
	P_{\mathrm{ D6 }} = \Gamma^{0123456} \sigma_1 \,,
	\notag\\
	P_{\mathrm{ D1 }} = \Gamma^{01} \sigma_1 \,, \qquad
	P_{\mathrm{ D3 }} = \Gamma^{0123} i\sigma_2 \,, &\qquad
	P_{\mathrm{ D5 }} = \Gamma^{012345} \sigma_1 \,.
	\notag
\end{align}

The projectors in M-theory are given by:
\begin{equation}
	P_{\mathrm{ P  }} = \Gamma^{01}\,, \qquad
	P_{\mathrm{ M2 }} = \Gamma^{012} \,, \qquad
	P_{\mathrm{ M5}} = \Gamma^{012345} \,, \qquad
	P_{\mathrm{ KKm}}^{\mathrm{ IIB}} = \Gamma^{0123456} \,.
\end{equation}



\bibliographystyle{JHEP}

\bibliography{microbib}

\providecommand{\href}[2]{#2}\begingroup\raggedright\begin{thebibliography}{10}

\bibitem{Strominger:1996sh}
A.~Strominger and C.~Vafa, \emph{{Microscopic Origin of the Bekenstein-Hawking
  Entropy}}, \href{https://doi.org/10.1016/0370-2693(96)00345-0}{\emph{Phys.
  Lett.} {\bfseries B379} (1996) 99}
  [\href{https://arxiv.org/abs/hep-th/9601029}{{\ttfamily hep-th/9601029}}].

\bibitem{Maldacena:1996ky}
J.~M. Maldacena, \emph{{Black holes in string theory}},
  \href{https://arxiv.org/abs/hep-th/9607235}{{\ttfamily hep-th/9607235}}.

\bibitem{Dijkgraaf:1996cv}
R.~Dijkgraaf, E.~P. Verlinde and H.~L. Verlinde, \emph{{BPS spectrum of the
  five-brane and black hole entropy}},
  \href{https://doi.org/10.1016/S0550-3213(96)00638-4}{\emph{Nucl. Phys. B}
  {\bfseries 486} (1997) 77}
  [\href{https://arxiv.org/abs/hep-th/9603126}{{\ttfamily hep-th/9603126}}].

\bibitem{Giusto:2004kj}
S.~Giusto and S.~D. Mathur, \emph{{Geometry of D1-D5-P bound states}},
  \href{https://doi.org/10.1016/j.nuclphysb.2005.09.037}{\emph{Nucl. Phys.}
  {\bfseries B729} (2005) 203}
  [\href{https://arxiv.org/abs/hep-th/0409067}{{\ttfamily hep-th/0409067}}].

\bibitem{Bena:2005va}
I.~Bena and N.~P. Warner, \emph{{Bubbling supertubes and foaming black holes}},
  \href{https://doi.org/10.1103/PhysRevD.74.066001}{\emph{Phys. Rev.}
  {\bfseries D74} (2006) 066001}
  [\href{https://arxiv.org/abs/hep-th/0505166}{{\ttfamily hep-th/0505166}}].

\bibitem{Berglund:2005vb}
P.~Berglund, E.~G. Gimon and T.~S. Levi, \emph{{Supergravity microstates for
  BPS black holes and black rings}},
  \href{https://doi.org/10.1088/1126-6708/2006/06/007}{\emph{JHEP} {\bfseries
  0606} (2006) 007} [\href{https://arxiv.org/abs/hep-th/0505167}{{\ttfamily
  hep-th/0505167}}].

\bibitem{Bena:2006is}
I.~Bena, C.-W. Wang and N.~P. Warner, \emph{{The foaming three-charge black
  hole}}, \href{https://doi.org/10.1103/PhysRevD.75.124026}{\emph{Phys. Rev.}
  {\bfseries D75} (2007) 124026}
  [\href{https://arxiv.org/abs/hep-th/0604110}{{\ttfamily hep-th/0604110}}].

\bibitem{Bena:2006kb}
I.~Bena, C.-W. Wang and N.~P. Warner, \emph{{Mergers and Typical Black Hole
  Microstates}},
  \href{https://doi.org/10.1088/1126-6708/2006/11/042}{\emph{JHEP} {\bfseries
  11} (2006) 042} [\href{https://arxiv.org/abs/hep-th/0608217}{{\ttfamily
  hep-th/0608217}}].

\bibitem{Bena:2007kg}
I.~Bena and N.~P. Warner, \emph{{Black holes, black rings and their
  microstates}}, \href{https://doi.org/10.1007/978-3-540-79523-0}{\emph{Lect.
  Notes Phys.} {\bfseries 755} (2008) 1}
  [\href{https://arxiv.org/abs/hep-th/0701216}{{\ttfamily hep-th/0701216}}].

\bibitem{Bena:2007qc}
I.~Bena, C.-W. Wang and N.~P. Warner, \emph{{Plumbing the Abyss: Black Ring
  Microstates}},
  \href{https://doi.org/10.1088/1126-6708/2008/07/019}{\emph{JHEP} {\bfseries
  07} (2008) 019} [\href{https://arxiv.org/abs/0706.3786}{{\ttfamily
  0706.3786}}].

\bibitem{Bena:2008wt}
I.~Bena, N.~Bobev and N.~P. Warner, \emph{{Spectral Flow, and the Spectrum of
  Multi-Center Solutions}},
  \href{https://doi.org/10.1103/PhysRevD.77.125025}{\emph{Phys. Rev.}
  {\bfseries D77} (2008) 125025}
  [\href{https://arxiv.org/abs/0803.1203}{{\ttfamily 0803.1203}}].

\bibitem{Bena:2010gg}
I.~Bena, N.~Bobev, S.~Giusto, C.~Ruef and N.~P. Warner, \emph{{An
  Infinite-Dimensional Family of Black-Hole Microstate Geometries}},
  \href{https://doi.org/10.1007/JHEP03(2011)022,
  10.1007/JHEP04(2011)059}{\emph{JHEP} {\bfseries 1103} (2011) 022}
  [\href{https://arxiv.org/abs/1006.3497}{{\ttfamily 1006.3497}}].

\bibitem{Bena:2011dd}
I.~Bena, S.~Giusto, M.~Shigemori and N.~P. Warner, \emph{{Supersymmetric
  Solutions in Six Dimensions: A Linear Structure}},
  \href{https://doi.org/10.1007/JHEP03(2012)084}{\emph{JHEP} {\bfseries 1203}
  (2012) 084} [\href{https://arxiv.org/abs/1110.2781}{{\ttfamily 1110.2781}}].

\bibitem{Bena:2015bea}
I.~Bena, S.~Giusto, R.~Russo, M.~Shigemori and N.~P. Warner, \emph{{Habemus
  Superstratum! A constructive proof of the existence of superstrata}},
  \href{https://doi.org/10.1007/JHEP05(2015)110}{\emph{JHEP} {\bfseries 05}
  (2015) 110} [\href{https://arxiv.org/abs/1503.01463}{{\ttfamily
  1503.01463}}].

\bibitem{Bena:2016ypk}
I.~Bena, S.~Giusto, E.~J. Martinec, R.~Russo, M.~Shigemori, D.~Turton et~al.,
  \emph{{Smooth horizonless geometries deep inside the black-hole regime}},
  \href{https://doi.org/10.1103/PhysRevLett.117.201601}{\emph{Phys. Rev. Lett.}
  {\bfseries 117} (2016) 201601}
  [\href{https://arxiv.org/abs/1607.03908}{{\ttfamily 1607.03908}}].

\bibitem{Bena:2017geu}
I.~Bena, E.~Martinec, D.~Turton and N.~P. Warner, \emph{{M-theory Superstrata
  and the MSW String}},
  \href{https://doi.org/10.1007/JHEP06(2017)137}{\emph{JHEP} {\bfseries 06}
  (2017) 137} [\href{https://arxiv.org/abs/1703.10171}{{\ttfamily
  1703.10171}}].

\bibitem{Bena:2017xbt}
I.~Bena, S.~Giusto, E.~J. Martinec, R.~Russo, M.~Shigemori, D.~Turton et~al.,
  \emph{{Asymptotically-flat supergravity solutions deep inside the black-hole
  regime}}, \href{https://doi.org/10.1007/JHEP02(2018)014}{\emph{JHEP}
  {\bfseries 02} (2018) 014}
  [\href{https://arxiv.org/abs/1711.10474}{{\ttfamily 1711.10474}}].

\bibitem{Bena:2018mpb}
I.~Bena, E.~J. Martinec, R.~Walker and N.~P. Warner, \emph{{Early Scrambling
  and Capped BTZ Geometries}},
  \href{https://doi.org/10.1007/JHEP04(2019)126}{\emph{JHEP} {\bfseries 04}
  (2019) 126} [\href{https://arxiv.org/abs/1812.05110}{{\ttfamily
  1812.05110}}].

\bibitem{Ceplak:2018pws}
N.~\v{C}eplak, R.~Russo and M.~Shigemori, \emph{{Supercharging Superstrata}},
  \href{https://doi.org/10.1007/JHEP03(2019)095}{\emph{JHEP} {\bfseries 03}
  (2019) 095} [\href{https://arxiv.org/abs/1812.08761}{{\ttfamily
  1812.08761}}].

\bibitem{Heidmann:2019zws}
P.~Heidmann and N.~P. Warner, \emph{{Superstratum Symbiosis}},
  \href{https://doi.org/10.1007/JHEP09(2019)059}{\emph{JHEP} {\bfseries 09}
  (2019) 059} [\href{https://arxiv.org/abs/1903.07631}{{\ttfamily
  1903.07631}}].

\bibitem{Heidmann:2019xrd}
P.~Heidmann, D.~R. Mayerson, R.~Walker and N.~P. Warner, \emph{{Holomorphic
  Waves of Black Hole Microstructure}},
  \href{https://doi.org/10.1007/JHEP02(2020)192}{\emph{JHEP} {\bfseries 02}
  (2020) 192} [\href{https://arxiv.org/abs/1910.10714}{{\ttfamily
  1910.10714}}].

\bibitem{Mayerson:2020tcl}
D.~R. Mayerson, R.~A. Walker and N.~P. Warner, \emph{{Microstate Geometries
  from Gauged Supergravity in Three Dimensions}},
  \href{https://doi.org/10.1007/JHEP10(2020)030}{\emph{JHEP} {\bfseries 10}
  (2020) 030} [\href{https://arxiv.org/abs/2004.13031}{{\ttfamily
  2004.13031}}].

\bibitem{Shigemori:2020yuo}
M.~Shigemori, \emph{{Superstrata}},
  \href{https://doi.org/10.1007/s10714-020-02698-8}{\emph{Gen. Rel. Grav.}
  {\bfseries 52} (2020) 51} [\href{https://arxiv.org/abs/2002.01592}{{\ttfamily
  2002.01592}}].

\bibitem{Bena:2020yii}
I.~Bena, F.~Eperon, P.~Heidmann and N.~P. Warner, \emph{{The Great Escape:
  Tunneling out of Microstate Geometries}},
  \href{https://doi.org/10.1007/JHEP04(2021)112}{\emph{JHEP} {\bfseries 04}
  (2021) 112} [\href{https://arxiv.org/abs/2005.11323}{{\ttfamily
  2005.11323}}].

\bibitem{Bena:2020iyw}
I.~Bena, A.~Houppe and N.~P. Warner, \emph{{Delaying the Inevitable: Tidal
  Disruption in Microstate Geometries}},
  \href{https://doi.org/10.1007/JHEP02(2021)103}{\emph{JHEP} {\bfseries 02}
  (2021) 103} [\href{https://arxiv.org/abs/2006.13939}{{\ttfamily
  2006.13939}}].

\bibitem{Giusto:2020mup}
S.~Giusto, M.~R. Hughes and R.~Russo, \emph{{The Regge limit of AdS$_3$
  holographic correlators}},
  \href{https://arxiv.org/abs/2007.12118}{{\ttfamily 2007.12118}}.

\bibitem{Houppe:2020oqp}
A.~Houppe and N.~P. Warner, \emph{{Supersymmetry and Superstrata in Three
  Dimensions}},  \href{https://arxiv.org/abs/2012.07850}{{\ttfamily
  2012.07850}}.

\bibitem{Ganchev:2021pgs}
B.~Ganchev, A.~Houppe and N.~Warner, \emph{{Q-Balls Meet Fuzzballs: Non-BPS
  Microstate Geometries}},  \href{https://arxiv.org/abs/2107.09677}{{\ttfamily
  2107.09677}}.

\bibitem{Ganchev:2021iwy}
B.~Ganchev, A.~Houppe and N.~P. Warner, \emph{{New Superstrata from
  Three-Dimensional Supergravity}},
  \href{https://arxiv.org/abs/2110.02961}{{\ttfamily 2110.02961}}.

\bibitem{Bianchi:2016bgx}
M.~Bianchi, J.~F. Morales and L.~Pieri, \emph{{Stringy origin of 4d black hole
  microstates}}, \href{https://doi.org/10.1007/JHEP06(2016)003}{\emph{JHEP}
  {\bfseries 06} (2016) 003}
  [\href{https://arxiv.org/abs/1603.05169}{{\ttfamily 1603.05169}}].

\bibitem{Bianchi:2017bxl}
M.~Bianchi, J.~F. Morales, L.~Pieri and N.~Zinnato, \emph{{More on microstate
  geometries of 4d black holes}},
  \href{https://doi.org/10.1007/JHEP05(2017)147}{\emph{JHEP} {\bfseries 05}
  (2017) 147} [\href{https://arxiv.org/abs/1701.05520}{{\ttfamily
  1701.05520}}].

\bibitem{Heidmann:2017cxt}
P.~Heidmann, \emph{{Four-center bubbled BPS solutions with a Gibbons-Hawking
  base}}, \href{https://doi.org/10.1007/JHEP10(2017)009}{\emph{JHEP} {\bfseries
  10} (2017) 009} [\href{https://arxiv.org/abs/1703.10095}{{\ttfamily
  1703.10095}}].

\bibitem{Bena:2017fvm}
I.~Bena, P.~Heidmann and P.~F. Ramirez, \emph{{A systematic construction of
  microstate geometries with low angular momentum}},
  \href{https://doi.org/10.1007/JHEP10(2017)217}{\emph{JHEP} {\bfseries 10}
  (2017) 217} [\href{https://arxiv.org/abs/1709.02812}{{\ttfamily
  1709.02812}}].

\bibitem{Avila:2017pwi}
J.~Avila, P.~F. Ramirez and A.~Ruiperez, \emph{{One Thousand and One Bubbles}},
  \href{https://doi.org/10.1007/JHEP01(2018)041}{\emph{JHEP} {\bfseries 01}
  (2018) 041} [\href{https://arxiv.org/abs/1709.03985}{{\ttfamily
  1709.03985}}].

\bibitem{Tyukov:2018ypq}
A.~Tyukov, R.~Walker and N.~P. Warner, \emph{{The Structure of BPS Equations
  for Ambi-polar Microstate Geometries}},
  \href{https://doi.org/10.1088/1361-6382/aaf133}{\emph{Class. Quant. Grav.}
  {\bfseries 36} (2019) 015021}
  [\href{https://arxiv.org/abs/1807.06596}{{\ttfamily 1807.06596}}].

\bibitem{Kanitscheider:2006zf}
I.~Kanitscheider, K.~Skenderis and M.~Taylor, \emph{{Holographic anatomy of
  fuzzballs}}, \href{https://doi.org/10.1088/1126-6708/2007/04/023}{\emph{JHEP}
  {\bfseries 04} (2007) 023}
  [\href{https://arxiv.org/abs/hep-th/0611171}{{\ttfamily hep-th/0611171}}].

\bibitem{Kanitscheider:2007wq}
I.~Kanitscheider, K.~Skenderis and M.~Taylor, \emph{{Fuzzballs with internal
  excitations}}, {\emph{JHEP} {\bfseries 06} (2007) 056}
  [\href{https://arxiv.org/abs/0704.0690}{{\ttfamily 0704.0690}}].

\bibitem{Taylor:2007hs}
M.~Taylor, \emph{{Matching of correlators in AdS(3) / CFT(2)}},
  \href{https://doi.org/10.1088/1126-6708/2008/06/010}{\emph{JHEP} {\bfseries
  06} (2008) 010} [\href{https://arxiv.org/abs/0709.1838}{{\ttfamily
  0709.1838}}].

\bibitem{Giusto:2015dfa}
S.~Giusto, E.~Moscato and R.~Russo, \emph{{AdS$_{3}$ holography for 1/4 and 1/8
  BPS geometries}}, \href{https://doi.org/10.1007/JHEP11(2015)004}{\emph{JHEP}
  {\bfseries 11} (2015) 004}
  [\href{https://arxiv.org/abs/1507.00945}{{\ttfamily 1507.00945}}].

\bibitem{Bombini:2017sge}
A.~Bombini, A.~Galliani, S.~Giusto, E.~Moscato and R.~Russo, \emph{{Unitary
  4-point correlators from classical geometries}},
  \href{https://arxiv.org/abs/1710.06820}{{\ttfamily 1710.06820}}.

\bibitem{Giusto:2019qig}
S.~Giusto, S.~Rawash and D.~Turton, \emph{{Ads$_{3}$ holography at dimension
  two}}, \href{https://doi.org/10.1007/JHEP07(2019)171}{\emph{JHEP} {\bfseries
  07} (2019) 171} [\href{https://arxiv.org/abs/1904.12880}{{\ttfamily
  1904.12880}}].

\bibitem{Tormo:2019yus}
J.~Garcia~i Tormo and M.~Taylor, \emph{{One point functions for black hole
  microstates}}, \href{https://doi.org/10.1007/s10714-019-2566-6}{\emph{Gen.
  Rel. Grav.} {\bfseries 51} (2019) 89}
  [\href{https://arxiv.org/abs/1904.10200}{{\ttfamily 1904.10200}}].

\bibitem{Bena:2019azk}
I.~Bena, P.~Heidmann, R.~Monten and N.~P. Warner, \emph{{Thermal Decay without
  Information Loss in Horizonless Microstate Geometries}},
  \href{https://doi.org/10.21468/SciPostPhys.7.5.063}{\emph{SciPost Phys.}
  {\bfseries 7} (2019) 063} [\href{https://arxiv.org/abs/1905.05194}{{\ttfamily
  1905.05194}}].

\bibitem{Rawash:2021pik}
S.~Rawash and D.~Turton, \emph{{Supercharged AdS${}_3$ Holography}},
  \href{https://arxiv.org/abs/2105.13046}{{\ttfamily 2105.13046}}.

\bibitem{Ganchev:2021ewa}
B.~Ganchev, S.~Giusto, A.~Houppe and R.~Russo, \emph{{AdS$_3$ holography for
  non-BPS geometries}},  \href{https://arxiv.org/abs/2112.03287}{{\ttfamily
  2112.03287}}.

\bibitem{Shigemori:2019orj}
M.~Shigemori, \emph{{Counting Superstrata}},
  \href{https://doi.org/10.1007/JHEP10(2019)017}{\emph{JHEP} {\bfseries 10}
  (2019) 017} [\href{https://arxiv.org/abs/1907.03878}{{\ttfamily
  1907.03878}}].

\bibitem{Mayerson:2020acj}
D.~R. Mayerson and M.~Shigemori, \emph{{Counting D1-D5-P microstates in
  supergravity}},
  \href{https://doi.org/10.21468/SciPostPhys.10.1.018}{\emph{SciPost Phys.}
  {\bfseries 10} (2021) 018}
  [\href{https://arxiv.org/abs/2010.04172}{{\ttfamily 2010.04172}}].

\bibitem{Seiberg:1997zk}
N.~Seiberg, \emph{{New theories in six-dimensions and matrix description of M
  theory on T**5 and T**5 / Z(2)}},
  \href{https://doi.org/10.1016/S0370-2693(97)00805-8}{\emph{Phys. Lett. B}
  {\bfseries 408} (1997) 98}
  [\href{https://arxiv.org/abs/hep-th/9705221}{{\ttfamily hep-th/9705221}}].

\bibitem{Kutasov:2001uf}
D.~Kutasov, \emph{{Introduction to little string theory}}, {\emph{ICTP Lect.
  Notes Ser.} {\bfseries 7} (2002) 165}.

\bibitem{Callan:1997kz}
C.~G. Callan and J.~M. Maldacena, \emph{{Brane death and dynamics from the
  Born-Infeld action}},
  \href{https://doi.org/10.1016/S0550-3213(97)00700-1}{\emph{Nucl. Phys. B}
  {\bfseries 513} (1998) 198}
  [\href{https://arxiv.org/abs/hep-th/9708147}{{\ttfamily hep-th/9708147}}].

\bibitem{Rychkov:2005ji}
V.~S. Rychkov, \emph{{D1-D5 black hole microstate counting from supergravity}},
  \href{https://doi.org/10.1088/1126-6708/2006/01/063}{\emph{JHEP} {\bfseries
  01} (2006) 063} [\href{https://arxiv.org/abs/hep-th/0512053}{{\ttfamily
  hep-th/0512053}}].

\bibitem{CabreraPalmer:2004asc}
B.~Cabrera~Palmer and D.~Marolf, \emph{{Counting supertubes}},
  \href{https://doi.org/10.1088/1126-6708/2004/06/028}{\emph{JHEP} {\bfseries
  06} (2004) 028} [\href{https://arxiv.org/abs/hep-th/0403025}{{\ttfamily
  hep-th/0403025}}].

\bibitem{Bena:2014qxa}
I.~Bena, M.~Shigemori and N.~P. Warner, \emph{{Black-Hole Entropy from
  Supergravity Superstrata States}},
  \href{https://doi.org/10.1007/JHEP10(2014)140}{\emph{JHEP} {\bfseries 1410}
  (2014) 140} [\href{https://arxiv.org/abs/1406.4506}{{\ttfamily 1406.4506}}].

\bibitem{Dabholkar:1995nc}
A.~Dabholkar, J.~P. Gauntlett, J.~A. Harvey and D.~Waldram, \emph{{Strings as
  Solitons \& Black Holes as Strings}},
  \href{https://doi.org/10.1016/0550-3213(96)00266-0}{\emph{Nucl. Phys.}
  {\bfseries B474} (1996) 85}
  [\href{https://arxiv.org/abs/hep-th/9511053}{{\ttfamily hep-th/9511053}}].

\bibitem{Bena:2022sge}
I.~Bena, N.~Ceplak, S.~Hampton, Y.~Li, D.~Toulikas and N.~P. Warner,
  \emph{{Resolving Black-Hole Microstructure with New Momentum Carriers}},
  \href{https://arxiv.org/abs/2202.08844}{{\ttfamily 2202.08844}}.

\bibitem{Mathur:2013nja}
S.~D. Mathur and D.~Turton, \emph{{Oscillating supertubes and neutral rotating
  black hole microstates}},
  \href{https://doi.org/10.1007/JHEP04(2014)072}{\emph{JHEP} {\bfseries 04}
  (2014) 072} [\href{https://arxiv.org/abs/1310.1354}{{\ttfamily 1310.1354}}].

\bibitem{Bena:2013ora}
I.~Bena, S.~F. Ross and N.~P. Warner, \emph{{On the Oscillation of Species}},
  \href{https://doi.org/10.1007/JHEP09(2014)113}{\emph{JHEP} {\bfseries 09}
  (2014) 113} [\href{https://arxiv.org/abs/1312.3635}{{\ttfamily 1312.3635}}].

\bibitem{Bena:2011uw}
I.~Bena, J.~de~Boer, M.~Shigemori and N.~P. Warner, \emph{{Double, Double
  Supertube Bubble}},
  \href{https://doi.org/10.1007/JHEP10(2011)116}{\emph{JHEP} {\bfseries 10}
  (2011) 116} [\href{https://arxiv.org/abs/1107.2650}{{\ttfamily 1107.2650}}].

\bibitem{Lunin:2001fv}
O.~Lunin and S.~D. Mathur, \emph{{Metric of the multiply wound rotating
  string}}, \href{https://doi.org/10.1016/S0550-3213(01)00321-2}{\emph{Nucl.
  Phys.} {\bfseries B610} (2001) 49}
  [\href{https://arxiv.org/abs/hep-th/0105136}{{\ttfamily hep-th/0105136}}].

\bibitem{Lunin:2002iz}
O.~Lunin, J.~M. Maldacena and L.~Maoz, \emph{{Gravity solutions for the D1-D5
  system with angular momentum}},
  \href{https://arxiv.org/abs/hep-th/0212210}{{\ttfamily hep-th/0212210}}.

\bibitem{Mateos:2001qs}
D.~Mateos and P.~K. Townsend, \emph{{Supertubes}},
  \href{https://doi.org/10.1103/PhysRevLett.87.011602}{\emph{Phys. Rev. Lett.}
  {\bfseries 87} (2001) 011602}
  [\href{https://arxiv.org/abs/hep-th/0103030}{{\ttfamily hep-th/0103030}}].

\bibitem{Emparan:2001ux}
R.~Emparan, D.~Mateos and P.~K. Townsend, \emph{{Supergravity supertubes}},
  {\emph{JHEP} {\bfseries 07} (2001) 011}
  [\href{https://arxiv.org/abs/hep-th/0106012}{{\ttfamily hep-th/0106012}}].

\bibitem{Bena:2008dw}
I.~Bena, N.~Bobev, C.~Ruef and N.~P. Warner, \emph{{Supertubes in Bubbling
  Backgrounds: Born-Infeld Meets Supergravity}},
  \href{https://doi.org/10.1088/1126-6708/2009/07/106}{\emph{JHEP} {\bfseries
  07} (2009) 106} [\href{https://arxiv.org/abs/0812.2942}{{\ttfamily
  0812.2942}}].

\bibitem{Constable:1999ac}
N.~R. Constable, R.~C. Myers and O.~Tafjord, \emph{{The Noncommutative bion
  core}}, \href{https://doi.org/10.1103/PhysRevD.61.106009}{\emph{Phys. Rev. D}
  {\bfseries 61} (2000) 106009}
  [\href{https://arxiv.org/abs/hep-th/9911136}{{\ttfamily hep-th/9911136}}].

\bibitem{Bena:2016oqr}
I.~Bena, J.~Bl\r{a}b\"ack, R.~Minasian and R.~Savelli, \emph{{There and back
  again: A T-brane's tale}},
  \href{https://doi.org/10.1007/JHEP11(2016)179}{\emph{JHEP} {\bfseries 11}
  (2016) 179} [\href{https://arxiv.org/abs/1608.01221}{{\ttfamily
  1608.01221}}].

\bibitem{Bates:2003vx}
B.~Bates and F.~Denef, \emph{{Exact solutions for supersymmetric stationary
  black hole composites}},
  \href{https://doi.org/10.1007/JHEP11(2011)127}{\emph{JHEP} {\bfseries 1111}
  (2011) 127} [\href{https://arxiv.org/abs/hep-th/0304094}{{\ttfamily
  hep-th/0304094}}].

\bibitem{Balasubramanian:2006gi}
V.~Balasubramanian, E.~G. Gimon and T.~S. Levi, \emph{{Four Dimensional Black
  Hole Microstates: From D-branes to Spacetime Foam}},
  \href{https://doi.org/10.1088/1126-6708/2008/01/056}{\emph{JHEP} {\bfseries
  0801} (2008) 056} [\href{https://arxiv.org/abs/hep-th/0606118}{{\ttfamily
  hep-th/0606118}}].

\bibitem{Horowitz:1993wt}
G.~T. Horowitz and D.~L. Welch, \emph{{Duality invariance of the Hawking
  temperature and entropy}},
  \href{https://doi.org/10.1103/PhysRevD.49.R590}{\emph{Phys. Rev. D}
  {\bfseries 49} (1994) 590}
  [\href{https://arxiv.org/abs/hep-th/9308077}{{\ttfamily hep-th/9308077}}].

\bibitem{Lin:2022rzw}
H.~W. Lin, J.~Maldacena, L.~Rozenberg and J.~Shan, \emph{{Holography for people
  with no time}},  \href{https://arxiv.org/abs/2207.00407}{{\ttfamily
  2207.00407}}.

\end{thebibliography}\endgroup

\end{document}